\documentclass[graybox]{svmult}

\usepackage{type1cm}        
\usepackage{makeidx}         
\usepackage{graphicx}        
\usepackage{multicol}        
\usepackage[bottom]{footmisc}

\usepackage{newtxtext}       %
\usepackage{newtxmath}       
\UseRawInputEncoding

\makeindex             

\begin{document}

\title*{Lorentz Symmetry Violation in String-Inspired Effective Modified Gravity Theories}
\author{Nick E. Mavromatos}
\institute{Nick E. Mavromatos \at National Technical University of Athens, School of Applied Mathematical and Physical Sciences, Department of Physics. Zografou Campus 157 80, Athens, Greece, and \\ King's College London, Department of Physics, Strand, London WC2R 2LS, UK. \\
}
%
%
\maketitle

\abstract*{Each chapter should be preceded by an abstract (no more than 200 words) that summarizes the content. The abstract will appear \textit{online} at \url{www.SpringerLink.com} and be available with unrestricted access. This allows unregistered users to read the abstract as a teaser for the complete chapter.
Please use the 'starred' version of the \texttt{abstract} command for typesetting the text of the online abstracts (cf. source file of this chapter template \texttt{abstract}) and include them with the source files of your manuscript. Use the plain \texttt{abstract} command if the abstract is also to appear in the printed version of the book.}

\abstract{We discuss situations under which Lorentz symmetry is violated in effective gravitational field theories that arise in the low-energy limit of strings. In particular, we discuss spontaneous violation of the symmetry by the ground state of the system.  
In the flat space-time limit, the effective theory of the broken Lorentz Symmetry 
acquires a form that belongs to the general framework of the so-called Standard Model Extension (SME) formalism. A brief review of this formalism is given before we proceed to describe a concrete example, where we discuss a Lorentz-symmetry-Violating (LV) string-inspired cosmological model. The model is a gravitational field theory coupled to matter, which contains torsion, arising from the fundamental degrees of freedom of the underlying string theory. The latter, under certain conditions which we shall specify, can acquire a LV condensate, and lead, via the appropriate equations of motion, to solutions that violate Lorentz and CPT (Charge-Parity-Time-Reversal) symmetry. The model is described by a specific form of an SME effective theory, with specific LV and CPT symmetry Violating coefficients, which depend on the microscopic parameters of the underlying string theory, and thus can be bounded by current-era phenomenology. }

\section{Lorentz- and CPT Symmetries in Particle Physics and Cosmology and their potential violation}
\label{sec:1}

Ignoring gravity, particle-physics theory and the respective phenomenology, as we understand them today, are based exclusively on Lorentz Symmetric formalisms. The Standard Model (SM) of Particle Physics, which is a mathematically consistent gauge field theory of the electromagnetic, weak and strong interactions, in flat space-time backgrounds, is a relativistic ({\it i.e} Lorentz invariant), unitary quantum field theory, with local, renormalizable interactions. As such, it satisfies the important CPT theorem~\cite{cptbook,bj}, proved independently by Schwinger~\cite{cpt1}, L\"uders~\cite{cpt2}, Pauli~\cite{cpt3}, Bell~\cite{cpt4} and Jost~\cite{cpt5}, which states that such field theories are described by Lagrangian densities that are invariant under the successive action (in any order) of the generators of the discrete symmetries of Charge Conjugation (C), Parity (or spatial reflexion symmetry) (P) and Time Reversal (T). Although it is often stated that Lorentz invariance violation is somewhat fundamental in inducing CPT violation~\cite{greenberg1,greenberg2}, nonetheless there have been objections to this statement, through explicit examples given in \cite{cptl1,cptl2}, which support the thesis that the aforementioned conditions for the validity of the CPT theorem, that is, locality, unitarity and Lorentz invariance, are truly independent, since, for instance, non-local but otherwise Lorentz invariant models could be explicitly constructed which violate CPT. Indeed, the proof of the theorem of \cite{greenberg1,greenberg2} necessitates well-defined time-ordered products and transfer (thus scattering) matrices, which exclude non-local or non-unitrary models, for which scattering matrices are not well defined. 

This CPT symmetry has important implications for particle physics in that it implies equality of masses $m$, lifetimes (or equivalently decay widths $\Gamma$), magnitude (with opposite sign) of electric charges $q^+=-q^-$, and magnetic dipole moments $g_m$, between particles (matter) and antiparticles (antimatter). The most stringent experimental bound between particle-antiparticle mass differences to date refers to the neutral-Kaon system, ${K^0, \overline K^0}$~\cite{PDG}:
\begin{align}\label{massdiff}
\frac{m^{K^0} - m^{\overline K^0}}{m^{K^0} + m^{\overline K^0} }  < 10^{-18}, \quad {\rm with} \quad  \frac{\Gamma(K^0) - \Gamma(\overline{K^0})}{\frac{1}{2}(\Gamma^{K^0} + \Gamma^{\overline K^0})} < 10^{-17}\,,
\end{align}
where $K^0$ is the neutral kaon,  $m^{K^0}$ its (rest) mass, and the overline above a symbol denotes a quantity referring to the corresponding antiparticle. For completeness, we mention that the most stringent current upper bounds in the differences between proton ($p$)-antiproton($\overline p$)  electric charges and electron ($e^-$)-positron($e^+$) magnetic dipole moments are~\cite{PDG}
\begin{align}\label{chmagmom}
q(p) + q(\overline p) < 10^{-21} \, e  \,,  \quad \frac{g_m(e^+) - g_m(e^-)}{\frac{1}{2}(g_m(e^+) - g_m(e^-))} < 2 \times 10^{-12}
\end{align}
where $e$ is the electron charge (at zero energy scale). 

For atoms, CPT invariance means that the anti-matter atoms will have identical spectra with the corresponding matter atoms. We mention at this stage that, since antihydrogen has been produced in the Laboratory~\cite{antiH1,antiH2,antiH3}, it provides, together with other man-made antimatter atoms, such as antirprotonic helium~\cite{antihel1,antihel2,antihel3}, a playground for additional tests of CPT invariance, at an atomic spectra level~\cite{antiHcpt,antiHcpt2}.

The implications of CPT invariance for the evolution and state of the Universe are also of immense importance. If CPT symmetry characterises a (yet elusive though) quantum theory of gravity, which is believed to describe the birth and dynamics of our Universe immediately after the Big Bang (i.e. at times after the Big Bang of order of the Planck time $t_{\rm Pl} \sim 5.4 \times 10^{-44}$~s), then matter and antimatter would have been generated in equal amounts in the early Universe. The dominance of matter over antimatter in the Cosmos, however, is overwhelming. Indeed, a plethora of observations, including cosmic microwave background (CMB) ones~\cite{planck}, as well measurements on the abundance of elements in the Universe (Big-Bang-Nucleosynthesis (BBN) data)~\cite{abund}, yield the following matter-antimatter asymmetry (or baryon-asymmetry in the universe (BAU), as it is alternatively called, due to the dominance of baryonic matter among the observable matter) :
\begin{equation}\label{bau}
\Delta n = \frac{n_{B}-n_{\overline{B}}}{n_{B}+n_{\overline{B}}}\sim\frac{n_{B}-n_{\overline{B}}}{s}=(8.4-8.9)\times10^{-11},
\end{equation}
at the early stages of the cosmic expansion, {\it i.e} at times $t \sim 10^{-6}$~s
and temperatures $T \gtrsim 1$~GeV. In the above expression,  $s$ denotes the
entropy density of the Universe, and $n_{B(\overline B)}$ the baryon (antibaryon) number densities. The above number essentialy implies the existence of one antiproton in $10^9$ protons in the Universe. 

In the framework of CPT-symmetric quantum field theories, in the absence of quantum gravity, which is a valid one at the regime of 
temperatures and times for which \eqref{bau} applies, one could generate such an asymmetry, provided the following conditions, postulated
by A.D. Sakharov~\cite{sakharov}, are met in the early Universe: 
\begin{enumerate}
\item
Baryon-number (B)-violating interactions that allow the generation of states with $B\neq 0$ starting from an initial
state with $B =0$;
\item
Interactions capable of distinguishing between matter and antimatter. Assuming CPT symmetry, this would
require violation of both C and CP;
\item
Since matter-antimatter asymmetry is impossible in chemical equilibrium, one also requires
some breakdown of chemical equilibrium during an epoch in the early Universe, otherwise any generated matter-antimatter asymmetry would be washed out by the reverse interaction.
\end{enumerate}
In the Standard Model of particle physics, which is a Lorentz and CPT invariant, unitary, local quantum field theory, the above conditions are met but {\it only qualitatively}. Indeed, Baryon number violation occurs due to quantum  chiral anomalies~\cite{shap1,shap2,shap3,gavel1,gavel2}, as a consequence of non-perturbative (instanton) effects of the electroweak gauge group SU(2), which lead to non-conservation of the 
chiral Baryon-number current $J^{\rm B\mu}$:
\begin{align}\label{anom}
\partial_\mu J^{\rm B \, \mu} \propto g^2 n_f \, {\rm Tr} (\mathbf F_{\mu\nu}\, \cdot  \mathbf F_{\mu\nu}) + {\rm Abelian~weak~hypercharge~U}_{\rm Y}{\rm (1)~terms}\,, 
\end{align}
where the Tr is over SU(2) gauge group indices, $g$ is the SU(2) coupling, $n_f$ is the flavour (generation) number, and $\mathbf F_{\mu\nu}$ is the field strength of the SU(2) gauge field. Due to the instanton effects, the system of the early Universe can tunnel through to a sector with non-zero baryon number from a state with a zero baryon number, and as a result there is induced B-number violation.\footnote{We remark that the chiral anomalies also induce the same amount of lepton number (L) violation, since $\partial_\mu J^{\rm B \, \mu} = \partial_\mu J^{\rm L \, \mu}$.}

Moreover, CP Violation (CPV) is known to charascterise the hadron sector of the Standard Model (it has been observed for the first time in the neutral Kaon system~\cite{Christenson}). 
However, the order of the observed CP violation in the quark sector of the Standard Model is several orders of magnitude smaller than the one required to produce the BAU \eqref{bau}. Moreover, CPV has still not been observed in the lepton sector. For these reasons, physicists attempt to extend the Standard Model in order to discover new sources of CPV that could explain the BAU according to Sakharov's conditions ({\it e.g.} supersymmetric models, extra dimensions, including strings, models with right-handed neutrinos {\it etc.}).

Minimal, not necessarily supersymmetric, Lorentz and CPT invariant field-theoretic extensions of the Standard Model, in (3+1)-dimensional space time, that could provide extra sources of CPV, are the ones augmented with massive right-handed neutrinos (RHN) in their spectra. Such models, with three species of heavy sterile Majorana RHN, might be used as providers - via the seesaw mechanism~\cite{seesaw1,seesaw2,seesaw3,seesaw4,seesaw5,seesaw6} - of light masses for (at least two of) the active neutrinos of the Standard Model, as required by the observed flavour oscillations~\cite{flavour}. In such models, in the early Universe, there is lepton asymmetry generation ({\it Leptogenesis}), through appropriate one-loop corrected decays of the RHN into Standard Model particles and antiparticles~\cite{leptofuya,leptogenesis}. In such processes,  which are CPT-conserving, the existence of a non-trivial CPV requires {\it more than one species} of Majorana neutrinos~\cite{leptogenesis} and at least one-loop corrections in the appropriate decay processes. These features lead to a difference in the respective CPV decay widths of the Majorana neutrino into standard-model particles and antiparticles, thus producing a Lepton-number (L) violation at an appropriate cosmological freeze-out point. We stress that tree-level decays and cases with only one species of Majorana neutrino lead to zero lepton asymmetry in CPT invariant models. 

Such lepton number asymmetry generation is then communicated to the baryon sector via equilibrated sphaleron processes~\cite{shap1,shap3}, which violate both B and L numbers, but preserve their difference B-L ({\it Baryogenesis}). This Baryogenesis through Leptogenesis mechanism is currently a very popular one for the generation of matter-antimatter asymmetry in the Universe~\cite{leptogenesis}.\footnote{In most leptogenesis scenarios, the RHN are superheavy, of masses close to the Grand Unification scale, $m_N \gtrsim 10^{14}$~GeV, as required by microscopic seesaw models (including supersymmetric ones). Nonetheless, 
there are also non-supersymmetric models~\cite{nuMSM1,nuMSM2}, termed the $\nu$Minimal Standard Model ($\nu$MSM), according to which the sterile Majorana neutrinos have masses spanning the range from a few GeV to $\mathcal O(10)$ keV, with the lightest having very weak couplings to the Standard Model sector, so that it has a life time longer than the lifetime of the Universe, and as such it can provide a candidate for (warm) dark matter. Baryogenesis mechanisms in this latter framework have been discussed in \cite{Canetti}.} 

Although there is a well established theoretical understanding of the above processes in the context of more or less conventional (i.e Lorentz and CPT invariant) particle physics models, nonetheless the lack of experimental evidence for the existence of additional sources (beyond the Standard Model) of CP violation, and right-handed neutrinos, may be a hint that some other, less conventional mechanism is in operation to explain the big question as to why {\it we exist}, that is, why there is this overwhelming dominance of matter over antimatter in our observable Universe. 
In this respect, a question arises as to whether the above processes of generating matter-antimatter asymmetry in the Universe could have a {\it geometric origin}, possibly due to quantum fluctuations of space time (quantum gravity), which are strong in the early Universe, and such that they violate Lorentz and CPT symmetry, leading to unconventional origins and processes for Lepto/Baryogenesis.

The theory of quantum gravity is still elusive, despite several theoretical attempts in the past and current centuries. One of the biggest questions associated with a consistent quantum theory of space time concerns the dynamical emergence of spacetime itself, and therefore the background independence of the theory. In some background-independent modern approaches to quantum gravity, e.g. the so-called {\it spin foam models}~\cite{foam1,foam2}, one starts from a rather abstract discrete set of states, which eventually condense to form dynamically the space-time continuum. Lorentz invariance in such models of quantum gravity, at least in the way we are familiar with from particle physics, may thus not be sacrosanct. We also mention at this stage, that 
in more conventional models, where a background space time is assumed, Wheeler has conjectured, many years ago, that microscopic black-hole and other topologically non trivial fluctuations of space time, may themselves give space time a ``foamy structure'' at Planck length scales~\cite{wheeler}, which may not respect Lorentz symmetry. Such structures may also hinder information from a low-energy observer, who conducts scattering experiments, which may lead to an effective decoherence of quantum matter in such space times. In such systems, the quantum operator corresponding to the generator of CPT symmetry may not be well-defined in the effective low-energy theory~\cite{wald}, leading to intrinsic CPT violation, which may have distinguishing features~\cite{omega,decoh} as compared to conventional violation of CPT symmetry, the latter occurring, for instance, as a result of violation of Lorentz invariance in effective local field theories~\cite{greenberg1}, in which the generator of CPT symmetry is well defined, but does not commute with the Hamiltonian operator of the system. 

In general, the idea that Lorentz (LV) and/or CPT Violation (CPTV) might characterise some approaches to quantum gravity, and their effective low-energy field theories, which may lead to interesting phenomenology, has gained attention in recent years, as a result of the increased sensitivity of experiments, especially cosmic multimessenger ones, to such violations. Although at present there is no experimental evidence for such violations, nonetheless the sensitivity of some experiments to some model parameters may reach Planck scale sensitivity, or even surpass it under some circumstances~\cite{LVpheno}, thus approaching the regime of quantum gravity. 

String theory~\cite{string}, which is one of the most successful to date attempts to unify gravity with the rest of the fundamental interactions in nature, but so far has been developed as a space-time background-dependent approach, 
is based perturbatively on well-defined scattering matrices, and as such, most of its effective low-energy field theories  so far are characterised by Lorentz and CPT invariance. There has also been an attempt to claim that non perturbative strings would also be characterised by some form of CPT invariance~\cite{dine}. Nonetheless, there is no rigorous proof that non-perturbative string theory is not characrterised by ground states which do violate Lorentz and/or CPT symmetries, leading to effective low-energy theories which are plagued by such violations. To the contrary, there are claims, supported by plausibility arguments, that such ground states do exist~\cite{samuel,koststrings1,koststrings2,lehnert}
in the landscape of (open) string vacua, thereby leading to the possibility of {\it spontaneous} Lorentz and CPT Violation in string theory
(although it must be said that the non-perturbative stability of such vacua has not been rigorously established, as yet). 

\begin{important}{Important}
It is the purpose of this book chapter to discuss another scenario for the spontaneous violation of Lorentz and CPT symmetry in the closed string sector, which in fact will also involve gravitational anomalies. As we shall see, the condensation of the corresponding anomaly currents, in the presence of primordial gravitational waves, will result in the {\it spontaneous breaking} of Lorentz and CPT symmetry, with far reaching consequences for unconventional Baryogenesis through Leptogenesis in the respective string-inspired cosmologies, but also for inflation~\cite{bms1,bms2,ms1,ms2}.
\end{important}

Before doing so, it is instructive to mention that a formalism for testing phenomenologically the predictions of local effective field theories with Lorentz and CPT Violation is the so-called Standard Model Extension (SME)~\cite{sme,sme2,sme3}, whose LV and CPTV parameters and their current bounds have been tabulated in \cite{smebounds}. An SME in the presence of gravitational backgrounds has also been formulated~\cite{smegrav}. In the next section we review briefly the SME formalism in flat space-time backgrounds, which will be of relevance to us here.

\section{The Standard-Model-Extension effective field theory formalism \label{sec:sme}}

The SME formalism~\cite{sme,sme2,sme3} assumes that the spontaneous breakdown of Lorentz and/or CPT symmetries arises in effective local interacting field theories, which are initially Lorentz and CPT invariant, respecting unitarity and locality, and as such can be expressed in terms of (an infinite in principle) set of local quantum field theory operators, involving general-coordinate invariant (and thus also locally, in space-time, Lorentz invariant, on account of the equivalence principle) products of tensorial field operators $\mathcal O_{\mu_1\mu_2\dots} ^{\rm SM} $ (with $\mu_i=0, \dots 3$, $i=1,2, \dots ,$  (3+1)-dimensional space-time indices), depending on the fields of the Standard Model, with field operators 
$\mathcal C^{\mu_1\mu_2\dots}$  involving fields beyond the Standard Model. The spontaneous breaking 
of Lorentz and/or CPT symmetries arises from condensation of the latter operators, which in this way obtain non-trivial constant vacuum expectation  values $\langle C^{\mu_1\mu_2\dots} \rangle = {\rm constant} \ne 0$ (``background tensors''):
\begin{align}
\mathcal O_{\mu_1\mu_2\dots} ^{\rm SM} \,\mathcal C^{\mu_1\mu_2\dots} \quad \stackrel{\rm condensation}{\Rightarrow} \quad \mathcal O_{\mu_1\mu_2\dots} ^{\rm SM} \, \langle \mathcal C^{\mu_1\mu_2\dots}\rangle \,, \quad \mu_i=0, \dots 3, \quad i=1,2,\dots\,.
\end{align}
The background tensors of (mass) dimension five and higher, are suppressed by appropriate inverse powers of the scale $\Lambda$ of new physics, beyond the Standard Model, up to which the effective SME is valid. 
A complete classification of dimension five LV and CPTV operators in the fermion, scalar (Higgs) and gauge sectors of the Standard Model extension in flat space-time (Minkowski) backgrounds, including interactions among these sectors, as well as modifications of the respective kinetic terms, has been provided in \cite{posp}. 

\begin{important} {Important} 
The following criteria for acceptable SME operators 
have been adopted (which also characterise operators of any dimension in the SME formalism):
\begin{enumerate}
\item The operators must be gauge invariant,
\item The operators must be Lorentz invariant, after contraction with a background tensor,
\item The operators must not be reduced to a total derivative (as this would imply that the respective operators would not contribute to the dynamics of the system),
\item The operators must not reduce to lower-dimension operators by the use of the Euler-Lagrange equations of motion,
\item The operators must  couple to an irreducible background tensor. 
\end{enumerate}

The SME formalism should be viewed as an effective field theory formalism, providing a framework to perform calculations that are associated with (spontaneous) violation of Lorentz (and CPT) symmetry which can be used in the respective tests. It is not meant to delve into the microscopic way by means of which the symmetry violating background condensate tensors $\langle C^{\mu_1\mu_2\dots} \rangle $ arise, if at all,  as this is a feature of the underlying ultraviolet (UV) complete theory of  quantum gravity. As already mentioned, the various background tensors constitute the parameters of the SME effective theory, whose experimental bounds from a plethora of diverse, terrestrial and extraterrestrial experiments/observations, including cosmological measurements, are tabulated and continuously updated in \cite{smebounds}.
\end{important}

For our purposes below, we shall restrict ourselves to the free fermion sector, and in particular to the lowest order (and simplest) SME effective Lagrangian~\cite{sme,sme2,sme3}:
\begin{align}\label{smef}
\mathcal L_{\rm eff}^{\rm SME, fermion} = \overline \psi(x) \Big(\frac{i}{2} \, \gamma^\mu \, \stackrel{\leftrightarrow}{\partial_\mu}
-  \mathcal M \Big) \psi (x), \quad \mathcal M =  m \mathbf 1  + a_\nu \, \gamma^\nu + b_\mu \, \gamma_5 \, \gamma^\nu ,
\end{align}
where $\psi(x)$ denote a generic fermion, that could be a chiral spinor or even a Majorana one (in the case of right-handed neutrinos), $m$ is its mass, the quantity $\mathbf 1$ denotes the identity in spinor space, and $\gamma_5=i\gamma^0\, \gamma^1\, \gamma^2\, \gamma^3$ is the chirality matrix. The coefficients $a_\mu$ and $b_\mu$ in the generalised mass term in \eqref{smef} are both LV and CPTV background vectors. 

\begin{exercise}
\begin{quotation} Consider the  Lagrangian of a Dirac fermion of mass $m$, coupled to electromagnetic, $A_\mu(x)$, and
axion (pseudoscalar) fields,  $b(x)$:
\begin{align}\label{Aapsi}
\mathcal L _{A,b,\psi}= -\frac{1}{4} F_{\mu\nu} \, F^{\mu\nu} 
+ \overline \psi (\Big(\frac{i}{2} \, \gamma^\mu \, \stackrel{\leftrightarrow}{\partial_\mu} 
- q \, A_\mu(x) \gamma^\mu  -m \mathbf 1 - i\, g_{\rm ch} \, \partial_\mu b(x) \, \gamma^\mu \, \gamma_5 \Big) \psi(x) 
\end{align}
where $e,g_{\rm ch} \in \mathbb  R$ denote the corresponding real couplings, and $F_{\mu\nu}$ is the Maxwell tensor. 

\begin{enumerate}
\item[\textbf{(i)}] First, show that under Lorentz transformations, including improper ones, i.e. parity P and time reversal T, the quantities 
$\overline \psi \, \gamma^\mu \psi$ are Lorentz invariant, while $\overline \psi \, \gamma^\mu \, \gamma_5 \psi$ transform as 
${\rm det}(\Lambda)\, \overline \psi \, \gamma^\mu \, \gamma_5 \psi$, where ${\rm det}(\Lambda)$ denote the determinant of the Lorentz transformation (including improper ones).

\item[\textbf{(ii)}] Then, by taking into account the way the vector $A_\mu(x)$ and pseudoscalar field $b(x)$ transform under such transformations, prove the Lorentz and CPT invariance of the Lagrangian \eqref{Aapsi}. 

\item[\textbf{(iii)}] Finally, by using the explicit transformations of the spinor {\it fields} under Parity (P), Charge conjugation (C) and Time reversal (T), and the corresponding transformations of the $A_\mu(x)$ and $b(x)$ fields, that you can find in standard quantum field theory books~\cite{cptbook,bj}, prove the CPT invariance of \eqref{Aapsi}, under any order of the combined application of  C, T, P. Pay special attention to argue that under the antiunitary T operation, the imaginary unit $i$ that appears in the Lagrangian transforms as $T \, i \, T^{-1}=-i$. 

\item[\textbf{(iv)}] Consider now the case of constant background vector fields $\langle A_\mu\rangle \equiv a_\mu $ = constant, 
and $\langle \partial_\mu b \rangle \equiv b_\mu $ = constant
whose values remain constant under Lorentz (proper or improper) transformations. Show that these terms violate both Lorentz and CPT invariance.
\end{enumerate} 
\end{quotation}

\end{exercise}

In the context of our string-inspired model~\cite{bms1,bms2,ms1,ms2}, we shall describe a mechanism for the dynamical generation of the LV and CPTV background $b_\mu$ in the effective field theory \eqref{smef}, through an appropriate condensate of gravitational waves in a string-inspired gravitational effective field theory with torsion and anomalies. In fact, as we shall see, the coefficient $b_\mu$ in this case will be associated with a condensation of the dual of a totally antisymmetric component of a torsion tensor in the (3+1)-dimensional spacetime arising from string compactitication. In our model, there is no generation of an $a_\mu$ background, so from now on we set this coefficient to zero.

We mention at this stage, that, phenomenologically, there are stringent bounds of the coefficient $b_\mu$ today, which amount to~\cite{smebounds}:
\begin{align}\label{coeff}
|b_0| < 0.2 ~\,\, {\rm eV}, \quad |b_i| < 10^{-31} ~\,\, {\rm GeV}.
\end{align}
We shall show that such bounds are quite naturally respected in our cosmological model, 
as a consequence of the cosmic (temperature) evolution of the LV and CPTV coefficients $b_\mu$, which are generated during the inflationary period, and remain undiluted in the radiation era~\cite{bms1,bms2,ms1,ms2}. 

As we shall discuss, these background vectors $b_\mu$ play an important r\^ole in inducing phenomenologically-relevant Leptogenesis in string-inspired  effective particle-physics models which involve RHN in their spectra~\cite{sarkar1,sarkar2,sarkar3,sarkar4,ms1}. It worths stressing already at this point that, unlike the conventional CPT and Lorentz invariant approaches~\cite{leptogenesis}, which require at least two species of (Majorana) RHN, and one-loop treatment for the respective decays, in order for the necessary CPV to be effective in producing the lepton asymmetry, this type of LV and CPTV Leptogenesis occurs at tree level, and one species of RHN suffices. As we shall see, it is the CPTV properties of the background vector $b_\mu$, in the presence of which the RHN decays into standard-model particles take place, that guarantee this. 

We stress that the association of $b_\mu$ to torsion provides a geometric origin to the cosmic matter-antimatter asymmetry generated in this way.
We also remark that, given the universal coupling of the torsion to all fermion species, including lepton and quarks of the Standard Model, such a LV and CPTV mechanism through torsion condensation, may also lead directly to matter-antimatter asymmetries (LV and CPTV direct baryogenesis~\cite{berto,popl}) in this Universe, without necessitating the presence of RHN, but we shall not explore these latter scenarios here.

\section{A string-inspired gravitational theory with torsion and anomalies}

We are now well motivated to start employing string theory considerations that will lead us to the effective  gravitational theory with the aforementioned LV and CPTV properties. To this end, we first remark that in closed string theory~\cite{string}, the bosonic massless gravitational mutliplet consists of a spin-zero (scalar) field, the dilaton $\Phi (x) $, a spin-two symmetric tensor field, the graviton, $g_{\mu\nu}(x) =g_{\nu\mu}(x)$, where $\mu,\nu$ are spacetime indices, and a spin-one antisymmetric tensor (or Kalb-Ramond (KR)) field
$B_{\mu\nu}(x)=-B_{\nu\mu}(x)$. In the phenomenologically-relevant case of superstrings, this multiplet belongs to the ground state of string theory, which is augmented by the (local) supersymmetry partners of these fields. In our approach we shall not discuss those partners, and concentrate only on the aforementioned bosonic fields. In the scenario of \cite{ms1}, which we follow as a prototype model for our discussion in this chapter,  we assume that supersymmetry is dynamically broken during a pre-inflationary epoch of the string-inspired Universe, and, as such, the supersymmetric partner fields acquire heavy masses, even close to the Planck scale in the scenarios advocated in \cite{ms1}, and therefore decouple from the low-energy spectrum, of relevance to our subsequent discussion.

We next remark that, in the framework of perturbative strings, the closed-string $\sigma$-model deformation describing the propagation of the string in a KR field background $B_{\mu\nu}$, is given by the world-sheet expression~\cite{string}:
\begin{align}\label{bu1}
 \Delta S^\sigma_B \equiv \int_{\Sigma^{(2)}} d^2 \sigma \, B_{\mu\nu}(X)\, \varepsilon^{AB} \partial_A X^\mu \, \partial_B X^\nu~,
 \,\, \mu, \nu =0, \dots 3,  \quad A,B =1, 2\,, 
\end{align}
where the integral is over the surface $\Sigma^{(2)}$, which corresponds to the string-tree-level world-sheet with the topology of a two-dimensional sphere $S^{(2)}$. For our purposes, such lowest genus world-sheet topologies suffice, given that string loop corrections, which would be associated with higher-genus world-sheet surfaces, are subdominant for weak string couplings we assume throughout; the indices $A,B=1,2$ are world-sheet indices, 
$\varepsilon_{AB}=-\varepsilon_{BA}$ is the world-sheet covariant Levi-Civita antisymmetric tensor, and $X^\mu$, $\mu=0, \dots 3$, are world-sheet fields, whose zero modes play the r\^ole of target-space coordinates. We have assumed that consistent string compactification~\cite{string} to (3+1) spacetime dimensions has taken place, whose details will not be discussed here.

It can be seen straightforwardly (using Stokes theorem, and taking into account that the spherical-like surface $\Sigma^{(2)}$ has no boundary) that the integrand in \eqref{bu1} is invariant under the following U(1) gauge transformation in target space (which is not related to electromagnetism):
\begin{align}\label{bu2}
B_{\mu\nu} \, \to \, B_{\mu\nu} + \partial_\mu \theta_\nu (X) - \partial_\nu \theta_\mu (X), 
\quad \mu, \nu =0, \dots 3, 
 \end{align}
where $\theta_\mu(X)$, $\mu=0, \dots 3,$ are  gauge parameters. 

\begin{exercise} 
\begin{quotation} Starting from the expression for the world-sheet deformation \eqref{bu1}, prove its invariance under the gauge transformation \eqref{bu2}.\end{quotation} 
\end{exercise}

This implies that the target-space effective action, which describes the low-energy limit of the string theory at hand, will be invariant under the U(1) gauge symmetry \eqref{bu2}, and, as such, it will depend only on the field strength 
of $B_{\mu\nu}$ :  
\begin {align}\label{HdB}
H_{\mu\nu\rho} = \partial_{[\mu} \, B_{\nu\rho]}\,, 
\end{align}
where the symbol 
$[\dots ]$ indicates total antisymmetrisation of the respective indices. 

However, in string theory~\cite{string}, 
cancellation between gauge and gravitational anomalies in the extra dimensional space requires the introduction of Green-Schwarz counterterms~\cite{gs}, which results in the modification of the field strength $H_{\mu\nu\rho}$  by the respective Chern-Simons (gravitational (``Lorentz'', L)  and gauge (Y)) anomalous terms :
\begin{align}\label{GSH}
\mathbf{{\mathcal H}} &= \mathbf{d} \mathbf{B} + \frac{\alpha^\prime}{8\, \kappa} \, \Big(\Omega_{\rm 3L} - \Omega_{\rm 3Y}\Big),  \nonumber \\
\Omega_{\rm 3L} &= \omega^a_{\,\,c} \wedge \mathbf{d} \omega^c_{\,\,a}
+ \frac{2}{3}  \omega^a_{\,\,c} \wedge  \omega^c_{\,\,d} \wedge \omega^d_{\,\,a},
\quad \Omega_{\rm 3Y} = \mathbf{A} \wedge  \mathbf{d} \mathbf{A} + \mathbf{A} \wedge \mathbf{A} \wedge \mathbf{A},
\end{align}
where we used differential form language, for notational convenience. In the above expression, $\mathcal H$ is a three-form, 
the symbol $\wedge$ denotes the exterior product among differential ($k,\ell$) forms (${\mathbf f}^{(k)} \wedge {\mathbf g}^{(\ell)} = (-1)^{k\, \ell}\, {\mathbf g}^{(\ell)} \wedge {\mathbf f}^{(k)}$), $\mathbf{A} \equiv \mathbf A_\mu \, dx^\mu$ denotes the Yang-Mills gauge field one form, and $\omega^a_{\,\,b} \equiv \omega^a_{\,\,\,\mu\,b}\, dx^\mu$ is the spin connection one form, with the Latin indices $a,b,c,d$ being tangent space (SO(1,3)) indices. The quantity $\alpha^\prime$ is the Regge slope $\alpha^\prime=M_s^{-2}$, where 
$M_s$ is the string mass scale, which is in general different from the reduced Planck scale 
in four space-time dimensions that enters the definition of the four-dimensional gravitational
constant $\kappa = \sqrt{8\pi\, {\rm G}} = M_{\rm Pl}^{-1}$, with $M_{\rm Pl} =  2.43 \times 10^{18}$~GeV  (we work in units of $\hbar=c=1$ throughout this work).

To lowest (zeroth) order in a perturbative expansion in powers of the Regge slope $\alpha^\prime$, {\it i.e.} to quadratic order in a derivative expansion, the low-energy effective four-dimensional action corresponding to the bosonic massless string multiplet, reads~\cite{string}:\footnote{In this work we follow the convention
for the signature of the metric $(+, -,-,- )$, and the definitions of the Riemann Curvature tensor
$R^\lambda_{\,\,\,\,\mu \nu \sigma} = \partial_\nu \, \Gamma^\lambda_{\,\,\mu\sigma} + \Gamma^\rho_{\,\, \mu\sigma} \, \Gamma^\lambda_{\,\, \rho\nu} - (\nu \leftrightarrow \sigma)$, the Ricci tensor $R_{\mu\nu} = R^\lambda_{\,\,\,\,\mu \lambda \nu}$, and the Ricci scalar $R = R_{\mu\nu}g^{\mu\nu}$.}
\begin{align}\label{sea}
S_B  =\; \int d^{4}x\sqrt{-g}\Big( \dfrac{1}{2\kappa^{2}} [-R + 2\, \partial_{\mu}\Phi\, \partial^{\mu}\Phi] - \frac{1}{6}\, e^{-4\Phi}\, {\mathcal H}_{\lambda\mu\nu}{\mathcal H}^{\lambda\mu\nu} + \dots \Big),
\end{align}
with the ellipses $\dots$ denoting higher-derivative terms, and possible dilaton potentials (arising from string loops or other mechanisms in effective string-inspired models, such as dilaton and 
non-critical-string cosmologies~\cite{aben,emn,Lahanas}, and pre-Big-Bang scenarios~\cite{prebb}.). 
The action \eqref{sea} can be found by either matching the corresponding (lowest order in derivatives) string scattering amplitudes with those obtained from the action \eqref{sea}, or by considering the world-sheet conformal invariance conditions ({\it i.e.} the vanishing of the corresponding Weyl-anomaly coefficients~\cite{string}) of the corresponding two-dimensional $\sigma$ model, which describes the propagation of strings in the backgrounds of $\Phi, g_{\mu\nu}$ and $B_{\mu\nu}$, and identify them 
with the corresponding equations of motion stemming from the effective action \eqref{sea}.\footnote{There are, of course, well-known ambiguities in such processes~\cite{amb1,amb2,amb3,amb4}, associated with local field redefinitions which leave the perturbative string scattering matrix invariant, according to the equivalence theorem of local quantum field theories~\cite{equiv1,equiv2}.  Such ambiguities, allow for instance, the string effective actions at quartic order in deriatives ($\mathcal O(\alpha^\prime$)) to be cast in the dilaton-Gauss-Bonnet combination~\cite{amb1}, which is gree from gravitational ghosts.}

In our approach we shall consider the dilaton field as fixed to an appropriate constant value, corresponding to minimisation of its potential, 
so that the string coupling $g_s=\exp (\Phi)$ is fixed to phenomenologically acceptable values~\cite{string}. Without loss of generality, then, we may set from now on $\Phi=0$. This is a self consistent procedure, as explained in \cite{bms2} (see also Exercise \ref{ex:dil} in section \ref{sec:rvm}) , which yields the $\Phi=0$ configuration as a solution for the dilaton equation  that acts as a constraint in this case. 

The torsion~\cite{torsion} interpretation of $\mathcal H_{\mu\nu\rho}$ arises by noticing that one can combine the quadratic in $\mathcal H_{\mu\nu\rho}$  terms of \eqref{sea} with the Einstein-Hilbert curvature scalar term $R$ in a generalised curvature scalar $\overline R(\overline \Gamma)$ with respect to a generalised connection, so that the action \eqref{sea}, with $\Phi=0$, is equivalent to the action:
\begin{align}\label{seator}
S_B  =\; \int d^{4}x\sqrt{-g}\, \dfrac{1}{2\kappa^{2}} \Big(-R(\overline \Gamma)  + \dots \Big),
\end{align}
where
\begin{align}\label{torsionconn}
{\overline \Gamma}_{\mu\nu}^{\rho} = \Gamma_{\mu\nu}^\rho + \frac{\kappa}{\sqrt{3}}\, {\mathcal H}_{\mu\nu}^\rho  \ne {\overline \Gamma}_{\nu\mu}^{\rho}\,\,
\end{align}
where $\Gamma_{\mu\nu}^\rho = \Gamma_{\nu\mu}^\rho$ is the torsion-free Christoffel symbol. Since the KR field strength satisfies 
\begin{align}\label{antisH}
\mathcal H^\mu_{\nu\rho} = -
\mathcal H^\mu_{\rho\nu}\,, 
\end{align}
it plays the r\^ole of contorsion~\cite{torsion}. This contorted geometry 
contains only a totally antisymmetric component of torsion~\cite{torsion}.\footnote{Using local field redefinition ambiguities~\cite{string,kaloper,amb1,amb2,amb3} one can extend the torsion interpretation of $\mathcal H$ to 
${\mathcal O}(\alpha^\prime)$ effective actions, which include fourth-order derivative terms.} 

\begin{exercise}
\begin{quotation} Prove the equivalence, up to total derivative terms, of the actions \eqref{sea} and \eqref{seator}, taking into account \eqref{torsionconn} and \eqref{antisH}.\end{quotation} 
\end{exercise}

The modification (\ref{GSH})
leads to the Bianchi identity (in differential form language)~\cite{string}
\begin{equation}\label{modbianchi}
\mathbf{d} \mathbf{{\mathcal H}} = \frac{\alpha^\prime}{8 \, \kappa} {\rm Tr} \Big(\mathbf{R} \wedge \mathbf{R} - \mathbf{F} \wedge \mathbf{F}\Big)
\end{equation}
where  $\mathbf{R}^a_{\,\,b} = \mathbf{d} \omega^a_{\,\,b} + \omega^a_{\,\,c} \wedge \omega^c_{\,\,b}$ is the curvature two form,  
$\mathbf{F} = \mathbf{d} \mathbf{A} + \mathbf{A} \wedge  \mathbf{A}$ is the Yang-Mills field-strength two form,  
and the trace (Tr) is over Lorentz- and gauge- group indices, respectively.
The non zero quantity on the right hand side  of \eqref{modbianchi} is the ``mixed (gauge and gravitational) quantum anomaly''~\cite{alvarez}. 
In the (more familiar) component form, the identity \eqref{modbianchi}, becomes:
\begin{align}\label{modbianchi2}
 \varepsilon_{abc}^{\;\;\;\;\;\;\;\;\;\mu}\, {\mathcal H}^{abc}_{\;\;\;\;\;\;\; ;\mu}
 =  \frac{\alpha^\prime}{32\, \kappa} \, \sqrt{-g}\, \Big(R_{\mu\nu\rho\sigma}\, \widetilde R^{\mu\nu\rho\sigma} -
F_{\mu\nu}\, \widetilde F^{\mu\nu}\Big) \equiv \sqrt{-g}\, {\mathcal G}(\omega, \mathbf{A}),
\end{align}
where the semicolon denotes gravitational covariant derivative with respect to the standard
Christoffel connection, and
\begin{align}\label{coveps}
\varepsilon_{\mu\nu\rho\sigma} = \sqrt{-g}\,  \epsilon_{\mu\nu\rho\sigma}\,, \quad \, \varepsilon^{\mu\nu\rho\sigma} =\frac{{\rm sgn}(g)}{\sqrt{-g}}\,  \epsilon^{\mu\nu\rho\sigma}\,, 
\end{align}
denote the gravitationally covariant Levi-Civita tensor densities, totally antisymmetric in their indices, with $\epsilon_{\mu\nu\rho\sigma}$ ($\epsilon_{0123} = +1$, {\emph etc.}) the Minkowski-space-time Levi-Civita totally antisymmetric symbol. 
The symbol
$\widetilde{(\dots)}$
over the curvature or gauge field strength tensors denotes the corresponding duals, defined as 
\begin{align}\label{duals}
\widetilde R^{\mu\nu\rho\sigma} \equiv \frac{1}{2} \varepsilon^{\mu\nu\lambda\pi} R^{\,\,\,\,\,\,\,\,\,\rho\sigma}_{\lambda\pi}\,, \, \quad  \widetilde F^{\mu\nu} \equiv \frac{1}{2} \varepsilon^{\mu\nu\rho\sigma}\, F_{\rho\sigma}\,, 
\end{align}
respectively.
The mixed-anomaly term is a total derivative
\begin{align}\label{pontryaginA}
&\sqrt{-g} \, \Big(R_{\mu\nu\rho\sigma}\, \widetilde R^{\mu\nu\rho\sigma} - F_{\mu\nu}\, \widetilde F^{\mu\nu} \Big) = \sqrt{-g} \, {\mathcal K}^\mu (\omega, \mathbf A)_{;\mu} = \partial_\mu \Big(\sqrt{-g} \, {\mathcal K}^\mu (\omega, \mathbf A) \Big) \nonumber \\
&= 2 \, \partial_\mu \Big[\epsilon^{\mu\nu\alpha\beta}\, \omega_\nu^{ab}\, \Big(\partial_\alpha \, \omega_{\beta ab} + \frac{2}{3}\, \omega_{\alpha a}^{\,\,\,\,\,\,\,c}\, \omega_{\beta cb}\Big) - 2 \epsilon^{\mu\nu\alpha\beta}\, \Big(A^i_\nu\, \partial_\alpha A_\beta^i + \frac{2}{3} \, f^{ijk} \, A_\nu^i\, A_\alpha^j \, A_\beta^k \Big)\Big],
\end{align}
where $i,j,k$ denote gauge group indices, with $f^{ijk}$ the gauge group structure constants.

\begin{exercise}
\begin{quotation} Using the definitions of the curvature and gauge field strength differential forms (in a shorthand notation, for brevity), $\mathbf R = d\omega + \omega \wedge \omega$ and 
$\mathbf F = d\mathbf A + \mathbf A \wedge \mathbf A$, in terms of the spin connection $\omega$ and the gauge field connection $\mathbf A$, respectively, prove Eq.~\eqref{modbianchi}, by taking the exterior derivative of the three form $\mathcal H$ in \eqref{GSH}.\end{quotation} 
\end{exercise}

\begin{important}{Important}
In our
four-dimensional cosmology~\cite{bms1,bms2,ms1,ms2} we shall {\it not} cancel the anomalies. 
In fact, we shall assume that only fields of the bosonic degrees of freedom of the massless gravitational string multiplet appear as external fields in the effective action describing the dynamics of the early Universe. Chiral fermionic and gauge matter are generated at the end of the inflationary period as we shall discuss later on. 
With the above assumptions, one may implement the Bianchi identity \eqref{modbianchi} as a {\it constraint} in a path-integral, via a pseudoscalar (axion-like) Lagrange multiplier field $b(x)$. After the ${\mathcal H}_{\mu\nu\rho}$ path-integration, then, one arrives at an effective action for the dynamics of the early epoch of the string-inspired Universe, which, upon the assumption of constant dilatons, contains only gravitons and the now dynamical field $b(x)$, canonically normalised, without potential, which corresponds to the massless string-model-independent gravitational (or KR)  axion field~\cite{kaloper,svrcek}: 
\begin{align}\label{sea4}
S^{\rm eff}_B &=\int d^{4}x\,\sqrt{-g}\Big[ -\dfrac{1}{2\kappa^{2}}\, R + \frac{1}{2}\, \partial_\mu b \, \partial^\mu b +   \sqrt{\frac{2}{3}}\,
\frac{\alpha^\prime}{96 \, \kappa} \, b(x) \, R_{\mu\nu\rho\sigma}\, \widetilde R^{\mu\nu\rho\sigma} + \dots \Big] \nonumber \\
&= \int d^{4}x\,\sqrt{-g}\Big[ -\dfrac{1}{2\kappa^{2}}\, R + \frac{1}{2}\, \partial_\mu b \, \partial^\mu b \Big] - 
\int d^{4}x \sqrt{\frac{2}{3}}\,
\frac{\alpha^\prime}{96 \, \kappa} \, b(x) \, R_{\mu\nu\rho\sigma}\, ^\ast R^{\mu\nu\rho\sigma} + \dots 
\nonumber \\
& =  \int d^{4}x\, \sqrt{-g}\Big[ -\dfrac{1}{2\kappa^{2}}\, R + \frac{1}{2}\, \partial_\mu b \, \partial^\mu b  -
 \sqrt{\frac{2}{3}}\,
\frac{\alpha^\prime}{96 \, \kappa} \, {\mathcal K}^\mu (\omega)\, \partial_\mu b(x)   + \dots \Big].
\end{align}
In passing from the first to the second line of \eqref{sea4} we have used the definitions \eqref{coveps} and that 
sgn(g)=-1. The symbol $^\ast R^{\mu\nu\rho\sigma}$ denotes the dual with respect to the flat-space-time Levi-Civita totally antisymmetric symbol $\epsilon^{\mu\nu\rho\sigma}$, with $\epsilon^{0123}=+1$, {\it etc.}:
\begin{align}\label{dualflat}
^\ast R^{\mu\nu\rho\sigma} \equiv \frac{1}{2} \epsilon^{\mu\nu\lambda\pi} R^{\,\,\,\,\,\,\,\,\,\rho\sigma}_{\lambda\pi}\,,
\end{align}
In the last line of \eqref{sea4} we have used \eqref{pontryaginA}, setting $\mathbf A=0$, and  performed appropriately the integration by parts, taking into account that fields and their first derivatives vanish at space-time infinity. The action \eqref{sea4} is nothing other than the action describing the Chern-Simons modification of general relativity in the presence of axion fields~\cite{jackiw,yunes}.
In fact, from this latter point of view, one may view this action as a generic Chern-Simons-modified-gravity action, beyond the specific context of string theory, in which case the coefficient of the Chern-Simons term  should be replaced by a generic real parameter:
\begin{align}\label{genericCS}
\sqrt{\frac{2}{3}} \frac{\alpha^\prime}{96\, \kappa} \quad \Rightarrow \quad\, \mathcal A_{\rm CS} \in \mathbb R\,, 
\end{align}
to be determined ``phenomenologically'' in various contexts ({\it e.g}., rotating black holes and wormholes, beyond string theory, as in ~\cite{YunBH1,YunBH2,chatz}). 
\end{important}

\begin{exercise}\label{bianchconstr}
\begin{quotation} Consider the path integral of the action \eqref{sea} with respect to the field $\mathcal H_{\mu\nu\rho}$, setting the dilaton $\Phi=0$:
\begin{align}\label{PIH} \mathcal Z_{\mathcal H} = \int \mathcal D \mathcal H \, \exp(i S_{\rm B}) \,,
\end{align}
where $\mathcal D \mathcal H $ denotes the appropriate path-integration measure. Insert the Bianchi-identity \eqref{modbianchi2}, in the absence of gauge fields ($\mathbf A=0$),  
as a $\delta$-functional constraint, $$\delta\Big(\varepsilon_{abc}^{\;\;\;\;\;\;\;\;\;\mu}\, {\mathcal H}^{abc}_{\;\;\;\;\;\;\; ;\mu}
 -  \frac{\alpha^\prime}{32\, \kappa} \, \sqrt{-g}\, \Big(R_{\mu\nu\rho\sigma}\, \widetilde R^{\mu\nu\rho\sigma}\Big)\Big),$$  in the integrand of \eqref{PIH}. By representing the $\delta (x)$ functional as an integral over a pseudoscalar Lagrange multiplier field, perform the $\mathcal H$-path integration, and normalise appropriately the Lagrange multiplier to link it to the field $b(x)$ appearing in the action \eqref{sea4}, with canonical kinetic term, thus mapping \eqref{PIH} to a path integral over $b(x)$ corresponding to the action \eqref{sea4}. Why the Lagrange multiplier field in this case has to be a pseudoscalar? \end{quotation} 
\end{exercise}

We note that classically, in (3+1) dimensional space-times,  the duality between $\mathcal H_{\mu\nu\rho}$ and $b(x)$ is provided by the relation (corresponding to saddle points of the $\mathcal H$ path-integral \eqref{PIH} after the $b$-representation of the Bianchi-constraint-\eqref{modbianchi2} $\delta$-functional)~\cite{kaloper,aben}
\begin{align}\label{dual}
-3\sqrt{2} \, \partial_\sigma b = \sqrt{-g} \, \epsilon_{\mu\nu\rho\sigma} \, \mathcal H^{\mu\nu\rho}.
\end{align}
The ellipses $\dots$ in \eqref{sea4} denote subdominant, for our purposes, higher derivative terms (in fact an infinity of them), but also 
other axions, arising from compactification in string theory~\cite{svrcek}, which 
have been discussed in \cite{ms2}, but will not be the focus of our present study. 
The reader should notice the presence of anomalous CP-violating couplings of the KR axion to gravitational anomalies in the action \eqref{sea4}. These will play an important r\^ole in inducing inflation in our string-inspired cosmology.

We note at this stage that, had we kept gauge fields in our early-universe cosmology as external fields, the KR axion field would also exhibit Lagrangian couplings of the form
$\propto b(x) {\rm Tr} \Big(\mathbf F_{\mu\nu} \, \widetilde{\mathbf F}^{\mu\nu}\Big)$. Such terms would not contribute to the stress tensor, being topological. 

\begin{exercise}
\begin{quotation} Prove that the contributions of the term $$b(x) {\rm Tr} \Big(\mathbf F_{\mu\nu} \, \widetilde{\mathbf F}^{\mu\nu}\Big)\,$$
where $\mathbf F_{\mu\nu}$ is the (non-Abelian, in general) gauge field strength, and $\widetilde{\mathbf F}^{\mu\nu}$ its dual (defined below Eq.~\eqref{modbianchi2}), 
to the stress-energy tensor of the theory vanish identically. \end{quotation} 
\end{exercise}

This needs to be contrasted with the 
gravitational anomaly terms in \eqref{sea4}, whose variation with respect to the metric field $g_{\mu\nu}$ yields non-trivial results~\cite{jackiw,yunes}:
\begin{align}\label{cottonvar}
\delta \Big[ \int d^4x \sqrt{-g} \, b \, R_{\mu\nu\rho\sigma}\, \widetilde R^{\mu\nu\rho\sigma} \Big] = 4 \int d^4x \sqrt{-g} \, {\mathcal C}^{\mu\nu}\, \delta g_{\mu\nu} = -
4 \int d^4x \sqrt{-g} \, {\mathcal C}_{\mu\nu}\, \delta g^{\mu\nu}\,, 
\end{align}
where 
\begin{align}\label{cottdef}
{\mathcal C}^{\mu\nu} \equiv  -\frac{1}{2}\, \Big[u_\sigma \, \Big( \varepsilon^{\sigma\mu\alpha\beta} R^\nu_{\, \, \beta;\alpha} +
\varepsilon^{\sigma\nu\alpha\beta} R^\mu_{\, \, \beta;\alpha}\Big) + u_{\sigma\tau} \, \Big(\widetilde R^{\tau\mu\sigma\nu} +
\widetilde R^{\tau\nu\sigma\mu} \Big)\Big]\,, 
\end{align}
is the (tracelss) Cotton tensor~\cite{jackiw},
\begin{align}\label{tracecott}
g^{\mu\nu} \, C_{\mu\nu} =0\,,
\end{align}
with $u_{\sigma} \equiv \partial_\sigma b = b_{;\sigma}, \,\,u_{\sigma\tau} \equiv  u_{\tau; \sigma} = b_{;\tau;\sigma}$.
Taking into account conservation properties of the Cotton tensor~\cite{jackiw}, 
\begin{align}\label{cottprop}
{\mathcal C}^{\mu\nu}_{\,\,\,\,\,\,\,;\,\mu} = \frac{1}{8}\, u^\nu \, R^{\alpha\beta\gamma\delta} \, \widetilde R_{\alpha\beta\gamma\delta}\,,
\end{align} we observe that the Einstein's equations stemming from \eqref{sea4} (and \eqref{genericCS}) read: 
\begin{align}\label{einsteincs}
R^{\mu\nu} - \frac{1}{2}\, g^{\mu\nu} \, R  - \mathcal A_{\rm CS} \,  {\mathcal C}^{\mu\nu} = \kappa^2 \, T^{\mu\nu}_{\rm matter},
\end{align}
where $T^{\mu\nu}_{\rm matter}$ denotes a matter stress tensor, which in our early-Universe cosmology 
includes only the KR axion-like field~\cite{bms1,bms2,ms1,ms2} 
\begin{align}\label{axionst}
T_{\mu\nu}^b= \partial_\mu b\, \partial_\nu b - \frac{1}{2} \Big(\partial_\alpha b \, \partial^\alpha b\Big)\, .
\end{align}
In more  general situations,  $T^{\mu\nu}_{\rm matter}$ contains all matter and radiation fields, but does {\it not} contain couplings to the curvature or derivatives of the metric tensor.

\begin{exercise} 
\begin{quotation} Prove the variational equation \eqref{cottonvar}  and the properties \eqref{tracecott} and \eqref{cottprop} of the Cotton tensor, using its definition \eqref{cottdef}.\end{quotation} 
\end{exercise}

\begin{important}{Important}

From the properties of the Einstein and Cotton tensors, stated above, we observe that the matter stress tensor is not conserved but it satisfies the conservation of an improved stress tensor in the form: 
\begin{align}\label{improved}
T^{\mu\nu}_{\rm improved \, ; \mu}  \equiv T^{\mu\nu}_{\rm matter \, ; \mu} + \mathcal A_{\rm CS}\,  {\mathcal C}^{\mu\nu}_{\,\,\,\,\,\,\,;\mu} =0\,.
\end{align} 
\begin{exercise}
\begin{quotation} Prove \eqref{improved}. \end{quotation} 
\end{exercise}

The presence of the Cotton tensor in this conservation equation  indicates exchange of energy between the KR axion field and the gravitational anomaly term, in a way consistent with diffeomorphism invariance and general covariance~\cite{bms1}. For Friedman-Lema$\hat{\rm i}$tre-Robertson-Walker (FLRW) geometries, the gravitational anomaly terms vanish, but this is not the case for (chiral) fluctuations about the FLRW background which violate CP invariance, for instance chiral gravitational-wave (GW) perturbations, as we discuss below.\end{important}

\section{Chiral gravitational-wave (quantum) fluctuations, anomaly condensates and Running-Vacuum-Model inflation \label{sec:condinfl}}

In the string-inspired cosmological model of \cite{bms1,ms1}, which is assumed to describe the dynamics of the early Universe, 
only fields from the massless gravitational string multiplet are assumed to appear in the effective gravitational action.
The generation of chiral fermionic and gauge matter occurs at the end of the inflationary era, as we shall discuss later on. 
For constant dilatons, we assumed so far, this implies that the effective Chern-Simons-modified gravity action \eqref{sea4} is the relevant one for a discussion of inflation in such a Universe. 
In the presence of (chiral) gravitational-wave (GW) quantum fluctuations of spacetime the anomaly terms are non trivial. In the literature there  have been essentially two ways of computing  the effects of GW on the gravitational anomaly terms: one is through Green's functions~\cite{stephon}, and the other through canonically quantised linearised gravity formalism~\cite{lyth}, which we adopt below as it seems closer to our spirit that the anomaly condensates are induced by quantum gravitational fluctuations.

\subsection{Chiral-gravitational-wave quantized perturbations} 

To this end, let one consider {\it quantised} tensor perturbations $h_{ij}(\eta, \vec x)$ in a FLRW  expanding Universe:
\begin{align}\label{dsmetr}
ds^2=a^2(\eta)\Big(d\eta^2 - (\delta_{ij} + 2 h_{ij}(\eta,\vec x)\Big)dx^idx^j, \quad i,j=1,2,3\, , 
\end{align}\label{hpert} 
where $a(\eta)$ is the scale factor of the FLRW Universe, and $\eta$ is the conformal time~\cite{kolb}, which is related in our approach to the Robertson-Walker time $t$ via 
\begin{align}\label{etat}
a(\eta) d\eta = + dt\,, 
\end{align}
(we consider the flow of both times in the same direction). The perturbations can be written in terms of their three-dimensional-space Fourier components, $h_p(\vec k, \eta)$, with $p=$\,Left ($L$) or Right ($R$), as~\cite{lyth}: 
\begin{align}
h_{ij}(\vec x, \eta) = \frac{\sqrt{2}}{M_{\rm Pl}}\, \int \frac{d^3 k}{(2\pi)^{3/2}} \, e^{i \vec k \cdot \vec x} \, 
\sum_{p={\rm L,\, R}}\, \epsilon^p_{ij}(\vec k)\, h_p(\vec k, \eta),
\end{align} 
with $\epsilon^p_{ij}(\vec k)$ the polarisation tensors, satisfying:
\begin{align}\label{pol}
k_i \epsilon^p_{ij}(\vec k) & =0, \quad \epsilon_{ij}^{p\,\star}(\vec k) \, \epsilon_{ij}^{p^\prime}(\vec k) = 2\, \delta_{pp^\prime}\,, \nonumber \\
\epsilon^{ilm}\, \epsilon_{ij}^{L\,\star}(\vec k) \, \epsilon_{jl}^{R}(\vec k) & = 
\epsilon^{ilm}\, \epsilon_{ij}^{R\,\star}(\vec k) \, \epsilon_{jl}^{L}(\vec k) = 0 \,,\nonumber \\
\epsilon^{ilm}\, \epsilon_{ij}^{L\,\star}(\vec k) \, \epsilon_{jl}^{L}(\vec k) & = -
\epsilon^{ilm}\, \epsilon_{ij}^{R\,\star}(\vec k) \, \epsilon_{jl}^{R}(\vec k) = -2i \frac{k_m}{|\vec k|}\,,
\end{align}
where the $\star$ denotes complex conjugation.

We quantise the tensor perturbations, assumed weak, by writing the tensor perturbation as an operator $\widehat h_{ij}(\vec k)$ in the Heisenberg picture, which implies that it satisfies the corresponding Einstein equation~\cite{lyth}:
\begin{align}\label{hquant}
\widehat h_{ij}(\vec x, \eta) &= \frac{\sqrt{2}}{M_{\rm Pl}}\, \int \frac{d^3 k}{(2\pi)^{3/2}} \,  
\sum_{p={L,\, R}}\, \Big(e^{i \vec k \cdot \vec x} \, \epsilon^p_{ij}(\vec k)\, \widehat h_p(\vec k, \eta) \,\Big)\,, \nonumber \\
\widehat h_p(\vec k, \eta) &= h_p(\vec k, \eta) \,\widehat a_p(\vec k) + h^\star_p(-\vec k, \eta) \,\widehat a^{\,\dagger}_p(-\vec k)\,,
\end{align} 
where the $\widehat{(\dots)}$ denotes a quantum operator,  and 
for the creation, $\widehat a^\dagger_p(\vec k)$, and annihilation operators, $\widehat a_p(\vec k)$, we have the canonical commutation relations
\begin{align}\label{cancomm}
\Big[ \widehat a_p(\vec k)\, , \, \widehat a^{\,\dagger}_{p^\prime}(\vec k^\prime)\Big] = \delta^{(3)}(\vec k - \vec k^\prime)\, 
\end{align} 
and the hermitian conjugate relation, with all others zero. The time-independent vacuum state $|0\rangle$  is defined by its annihilation by $\widehat a_p(\vec k)$, {\it i.e.} 
\begin{align}\label{vac}
\widehat a_p(\vec k) |0\rangle =0\,.
\end{align}  
Assuming weak tensor perturbations $h_{ij}$, \eqref{dsmetr}, we may show that, up to second order in such perturbations, the gravitational Chern Simons term assumes the form  
\begin{align}\label{rrtilde}
R_{\mu\nu\rho\sigma} \, ^{\ast} R^{\mu\nu\rho\sigma} = & -\frac{8}{a(\eta)^4} \epsilon^{ijk} \, \Big(\frac{\partial^2}{\partial x^l \, \partial \eta}
\, h_{jm} \, \frac{\partial^2}{\partial x^m \, \partial x^i} \, h_{kl} - \frac{\partial^2}{\partial x^l \, \partial \eta}
\, h_{jm} \, \frac{\partial^2}{\partial x^l \, \partial x^i} \, h_{km} \nonumber \\&+ \frac{\partial^2}{\partial \eta^2}
\, h_{jl} \, \frac{\partial^2}{\partial x^i \, \partial \eta} \, h_{lk}\Big),
\end{align} 
where $\epsilon^{ijk}$, $i,j,k =1,2,3$ spatial indices, is the totally antisymmetric symbol in Euclidean three-dimensional space. We remind the reader that the dual tensor $^\ast R^{\mu\nu\rho\sigma} $ is defined ({\it cf.} \eqref{dualflat}) with respect to the flat space-time Levi-Civita symbol  $\epsilon^{\mu\nu\rho\sigma}$, with $\epsilon^{0ijk}=+1 \equiv \epsilon^{ijk}$, $i,j,k=1,2,3$.

\begin{exercise}
\begin{quotation} Starting from the metric \eqref{dsmetr}, prove \eqref{rrtilde}, up to second order in weak tensor perturbations $h_{ij}$.
\end{quotation} 
\end{exercise} 

\begin{important}{Important}
The classical quantity \eqref{rrtilde} becomes a Chern-Simons operator $\widehat{R_{\mu\nu\rho\sigma} \, \widetilde R^{\mu\nu\rho\sigma}}$
upon replacing $h_{ij}(\eta,\vec x)$ by the corresponding quantum operators $\widehat h_{ij}(\eta,\vec x)$
\eqref{hquant}. Using \eqref{cancomm} and \eqref{vac}, one can show~\cite{lyth} that the vacuum expectation value of the Chern-Simons operator 
\begin{align}\label{CSop}
\langle 0 | \widehat{{ R_{\mu\nu\rho\sigma} \, ^\ast R^{\mu\nu\rho\sigma}}}|0\rangle = &\frac{16}{a(\eta)^4\, M_{\rm Pl}^2}\, 
\int \frac{d^3k}{(2\pi)^3} \Big[k^2 h_L^\star (k, \eta) \, h_L^\prime (k, \eta) - k^2 \, h^\star_R(k, \eta) \, h^\prime_R(k, \eta) 
\nonumber \\&- 
h_L^{\star\, \prime} (k, \eta)\, h_L^{\prime\prime}(k, \eta) +   h_R^{\star\, \prime} (k, \eta)\, h_R^{\prime\prime}(k, \eta)\Big]\, ,
\end{align}
where $k \equiv |\vec k|$, and the prime denotes derivative with respect to the conformal time $\eta$. Notice that the vacuum expectation value \eqref{CSop} vanishes in a Left-Right symmetric situation, therefore the result is only non zero when there is gravitational birefringence, {\it i.e.} chirality (in the sense of differences between left-right GW perturbations).
\end{important}

\begin{exercise} 
\begin{quotation}  Inserting \eqref{hquant} into \eqref{rrtilde}, 
and using canonical quantization properties, \eqref{cancomm} and \eqref{vac}, prove the result \eqref{CSop} for the vacuum expectation value of the gravitational Chern Simons operator.
\end{quotation} 
\end{exercise}
The physical momenta $k/(a(\eta))$ of the graviton modes should be cut-off at an UltraViolet (UV) scale $\mu$, which means that terms  in the ultraviolet regime dominate the integrals. It is at this point that to have a full understanding of the condensate one needs the UV complete theory of quantum gravity, such as the full string theory in this case. 

In \cite{stephon}, it was assumed that in the evolution equation of the graviton modes one can keep only up to second order derivative terms. Although, as correctly remarked in \cite{lyth}, this is far from a satisfactory treatment within a quantum gravity regime, as required by the fact that the dominant part of the Fourier integration is near the UV cut-off, where quantum gravity is fully operational, nonetheless, for our purposes of discussing qualitatively the effects of gravitational anomaly condensates on inducing a running vacuum inflation, this will suffice~\cite{bms1,bms2,ms1,ms2}, in view of the slow-roll of the weak axion KR field  that characterises our cosmology model, as we shall discuss below. 

Making this assumption, it can be easily seen that,
in an {\it inflationary (de Sitter-like) space-time background} of interest to us here, with an approximately constant Hubble parameter $H \simeq$ constant,  the normalised solutions of the Einstein equations \eqref{einsteincs}, in the presence of a weak anomaly term, assume the form:
\begin{align}\label{gravmodes}
h_{p=L,R} (k, \eta) \sim \exp(-i\, k \eta) \, \exp(\pm k \,\Theta (\eta-\eta_0))\, ,  \quad \Theta \equiv \frac{4}{a^2(\eta)} \, \Big(f^{\prime\prime}(b) + a(\eta) \, H\, f^\prime(b)\Big).
\end{align}
where
\begin{align}\label{deff}
f(b) \equiv \, \sqrt{\frac{2}{3}}\,\frac{1}{24}\, \frac{b(\eta)}{M_s^2\, M_{\rm Pl}} = \frac{3.4 \times 10^{-2}}{M_s^2\, M_{\rm Pl}}\, b(\eta) \,, 
\end{align}
with $\eta_0$ signalling the beginning of the inflationary phase in conformal-time-$\eta$ coordinates. The $\pm$ in the exponent of the second factor on the right-hand side of \eqref{gravmodes} refer to left (L), right (R) movers respectively. 
In arriving at \eqref{gravmodes} we ignored the ``matter'' contributions to the Einstein equations, as these are associated with the slowly rolling stiff-matter KR axions in our context (see below), and thus yield subleading terms, {\it quadratic} in $\dot b$~\cite{bms1,ms1} (the dot denotes derivative with respect to the cosmic Robertson-Walker time $t$, which is related to the conformal time $\eta$ via \eqref{etat}). Thus, the quantity $\Theta$ is assumed weak in our approach, $|\Theta | \ll 1$, due to the slow-roll assumption for the KR axion field, which $ \Theta$ depends upon ({\it cf.} \eqref{gravmodes},\eqref{deff}). In our approximations below, therefore, we keep only terms linear in $\Theta$.

On substituting \eqref{gravmodes} and \eqref{deff} in \eqref{CSop}, then, and performing the Fourier integrations, up to an UV  cutoff $\mu$~\cite{stephon,lyth}, such that $ k \le a(\eta)\, \mu $, we obtain to first order in $\Theta$:
~\cite{lyth}:
\begin{align}\label{rrstar}
\langle 0 | \widehat{{ R_{\mu\nu\rho\sigma} \, ^\ast R^{\mu\nu\rho\sigma}}}|0\rangle = \frac{4\, H^2}{\pi^2\, M_{\rm Pl}^2} \, \Theta \, \mu^4~. 
\end{align}
Passing from a conformal to Robertson-Walker cosmic time, we write $\Theta = 4 \ddot f + 8 \, H\, \dot f$. In our case $\ddot f \ll H\, \dot f $ as a consequence of the slow roll nature of the KR axion, which arises dynamically in a self consistent way, upon formation of gravitational anomaly condensates~\cite{bms1,bms2,ms1,ms2}, to be reviewed below. 
Thus, for our purposes, we may ignore the $\ddot f$ terms, and approximate:
\begin{align}\label{approxTheta}
\Theta \simeq \, 0.27\, \frac{H}{M_s^2\, M_{\rm Pl}}\, \dot b(t)\,,  
\end{align}
Hence, \eqref{rrstar} yields
\begin{align}\label{rrstar2}
\langle 0 | \widehat{{ R_{\mu\nu\rho\sigma} \, ^\star R^{\mu\nu\rho\sigma}}}|0\rangle \simeq \frac{1.1}{\pi^2} \, 
\Big(\frac{H}{M_{\rm Pl}}\Big)^3 \, \mu^4\, \frac{\dot b(t)}{M_s^2}~. 
\end{align}
In the specific context of string theory, it is reasonable to define the effective theory below the string mass scale, $M_s$, which therefore should act as an UV cutoff, hence we can identify $\mu \sim M_s$. ~On the other hand, in generic Chern-Simons modified gravity theories, the string scale enters only in the coefficient of the Chern-Simons anomaly term $\mathcal A_{\rm CS}$ \eqref{genericCS}, which is viewed as a phenomenological parameter. In such a context, it is natural to assume that the UV cut-off scale of the graviton modes is the Planck scale 
$\mu \sim M_{\rm Pl}$, while the string scale (and thus the magnitude of the coefficient $\mathcal A_{\rm CS}$) can be determined phenomenologically by discussing conditions for the formation of the 
anomaly condensate~\cite{bms2}, which we do in the next subsection. The reader should recall that our main aim in this work is to demonstrate that the formation of anomaly condensates will connect this Chern-Simons gravitational theory with the (Lorentz and CPT Violating) Standard Model Extension framework~\cite{sme,sme2}, described in section \ref{sec:sme}.

\subsection{Gravitational-Anomaly condensates and spontaneous violation of Lorentz symmetry
\label{sec:cond}}

In \cite{bms1,ms1} we have discussed the possibility of forming a condensate of the gravitational anomaly, in case there is a {\it macroscopic number} of sources of GW, with constructive interference. Below we shall clarify in some detail how we envisage the appearance of such a condensate, and what is its connection with the vacuum expectation value of the Chern Simons term operator \eqref{CSop} during inflation. It is understood that our approach will be phenomenological, providing only plausibility arguments for the condensate formation. The microscopic treatment requires a complete (non-perturbative) understanding of the underlying string theory or, more general, the UV complete theory of quantum gravity that characterises the effective theory, should one view the action \eqref{sea4} as a generic Chern-Simons modified gravity model, with a phenomenological Chern-Simons coefficient $\mathcal A_{\rm CS}$ \eqref{genericCS}.

To this end, we assume that the condensate is created by the collective effects of a (time-dependent) number $\mathcal N(t)$ of sources of GW per unit volume  in the expanding Universe (in this notation, the total number of sources is given by $N= \int d^4x \mathcal N (t) = \int d^4 x \sqrt{-g} \frac{\mathcal N(t)}{\sqrt{-g}}$, with $n_\star \equiv  \frac{\mathcal N(t)}{\sqrt{-g}}$ the proper number density of sources). In this case, the induced tensor perturbations in \eqref{dsmetr} are replaced by the sum
\begin{align}\label{sumdsmetr}
h_{ij}(n, \vec x) \quad \Rightarrow \quad \sum_{I=1}^{\mathcal N(t)} h^{I}_{ij}(\eta, \vec x) , 
\end{align}
expressing the collective effect of sources, where the index $I$ labels the source that produces a specific GW perturbation. In this effective, ``phenomenological'' approach, each individual $h_{ij}^{(I)}$ satisfies an Einstein equation of motion, but the collective metric 
induced under \eqref{sumdsmetr} does not, as a result of the dynamical (time dependent) nature of $\mathcal N(t)$, 
given that the dynamics of the formation of sources can only be dealt with in a full theory of UV quantum gravity, such as string theory {\it etc.}.\footnote{Indeed, one may  envisage that the primordial sources of GW appear dynamically 
as excitations of the ground state of the full quantum gravity system, and span the whole range, from non-spherically collapsing domain walls in a pre-inflationary epoch, to merging primordial black holes~\cite{ms1}, which are themselves created from (quantum) gravitational vacuum perturbations.} Nonetheless, we may proceed in a rather agnostic, phenomenological approach, and quantize each individual $h_{ij}^I$ by means of replacing it simply with operators $\widehat h_{ij}^I(\eta, \vec x)$ \eqref{hquant}, with the creation and annihilation operators, $\widehat a^{I\, \dagger}(\vec k), \, \widehat a^{I}(\vec k)$, respectively, now carrying a ``source'' index $I$. For the vacuum we demand
\begin{align}\label{vacN}
a^I(\vec k) |0\rangle =0, \quad {\rm and} \quad 
a^I(\vec k) \, a^{J\, \dagger}(\vec k^\prime) \, |0\rangle = \delta_{IJ}\, \delta^{(3)}(\vec k-\vec k^\prime) |0\rangle , \quad I,\,J= 1, \dots \mathcal N(t)\,,
\end{align}
where $\delta_{IJ}$ denotes a Kronecker delta.
It is then immediately seen, that, upon assuming that the dominant GW perturbations $h_p^I (\vec k, \eta)$ 
coming from these set of sources 
have all the same magnitude, so that the index $I$ can be omitted from the corresponding expression, we arrive at  
the analogue of \eqref{CSop} in this multi-source case:
\begin{align}\label{CSopN}
&\langle 0 | \widehat{{ R_{\mu\nu\rho\sigma} \, ^\ast R^{\mu\nu\rho\sigma}}}|0\rangle_{\mathcal N}  =  \mathcal N(t) \frac{16}{a^4\, M_{\rm Pl}^2}\, 
\int \frac{d^3k}{(2\pi)^3} \Big[k^2 h_L^\star (k, \eta) \, h_L^\prime (k, \eta)\nonumber \\ & - k^2 \, h^\star_R(k, \eta) \, h^\prime_R(k, \eta) 
- {h_L^\prime}^\star (k, \eta)\, h_L^{\prime\prime}(k, \eta) +   {h_R^\prime}^\star (k, \eta)\, h_R^{\prime\prime}(k, \eta)\Big]\, ,
\end{align}
that is, the collective effect is represented by a simple multiplication of the right-hand side of \eqref{CSop} by the number of sources. We stress again this is a plausible, but effective description, valid for weak GW perturbations from the various sources. This result, however, allows us now to represent the gravitational-anomaly condensate as
\begin{align}\label{condensateN}
\langle R_{\mu\nu\rho\sigma} \, \widetilde R^{\mu\nu\rho\sigma} \rangle_{\rm condensate\, \mathcal N} 
=\frac{1}{\sqrt{-g}}  \, 
\langle 0 | \widehat{{ R_{\mu\nu\rho\sigma} \, ^\ast R^{\mu\nu\rho\sigma}}}|0\rangle_{\mathcal N} \,,
\end{align} 
to quadratic order in the weak GW perturbations, where $\sqrt{-g}$ is the de Sitter {\it unperturbed} background metric. In arriving at \eqref{condensateN} we used \eqref{coveps} and \eqref{duals}. Using the estimate \eqref{rrstar2}, we can thus 
estimate the magnitude of the gravitational anomaly condensate, induced by a macroscopic number of sources 
$\mathcal N(t)$, as:
 \begin{align}\label{condensateN2}
\langle R_{\mu\nu\rho\sigma} \, \widetilde R^{\mu\nu\rho\sigma} \rangle_{\rm condensate\, \mathcal N} 
=\frac{\mathcal N(t)}{\sqrt{-g}}  \, \frac{1.1}{\pi^2} \, 
\Big(\frac{H}{M_{\rm Pl}}\Big)^3 \, \mu^4\, \frac{\dot b(t)}{M_s^{2}} \equiv n_\star \, \frac{1.1}{\pi^2} \, 
\Big(\frac{H}{M_{\rm Pl}}\Big)^3 \, \mu^4\, \frac{\dot b(t)}{M_s^{2}}~.
\end{align} 
Th reader should recall that in the above expression $n_\star \equiv \frac{\mathcal N(t)}{\sqrt{-g}} $ denotes the number density (over the proper de Sitter volume) 
of the sources. Without loss of generality, we may take this density to be (approximately) time independent during inflation. 

From the anomaly equation \eqref{pontryaginA}, which expresses the gravitational Chern-Simons term \eqref{condensateN2} as a divergence of an anomaly current, and assuming isotropy and homogeneity of the background space time (which we can justify 
microscopically in our framework through pre-inflationary epochs~\cite{ms1,ms2}) we may write, to leading order 
in GW pertubations during inflation~\cite{bms1,bms2,ms1}: 
\begin{align}\label{anomevol}
\langle \mathcal K^\mu_{\,\,\, \,;\,\mu} \rangle_{\rm condensate \mathcal N} \simeq  \frac{d}{dt}  <\mathcal K^0> + 3 \, H\, <\mathcal K^0>  \,
\simeq \, n_\star \,  \frac{1.1}{\pi^2} \, 
\Big(\frac{H}{M_{\rm Pl}}\Big)^3 \, \mu^4\, \frac{\dot b(t)}{M_s^2} \, , 
\end{align}
where $<\mathcal K^0>$ denotes the (dominant) average temporal component of the anomaly current in the de-Sitter background.

We next observe that from the Euler-Lagrange equations of the KR axion stemming from \eqref{sea4}, one obtains:
\begin{align}\label{krbsol}
\frac{1}{\sqrt{-g}} \, \partial_\mu \Big( \sqrt{-g} [\partial^\mu b  - \mathcal A_{\rm CS} \, \mathcal K^\mu ]\Big) =0\, ,
\end{align}
which for isotropic and homogeneous cosmological space times, leads to a solution (under the assumption of the formation of a condensate)
\begin{align}\label{LVsol}
\dot b = \mathcal A_{\rm CS} \, <\mathcal K^0> \,,
\end{align}  
where in the case of strings, the Chern-Simons coefficient $\mathcal A_{\rm CS} $ is defined in \eqref{genericCS} (see also \eqref{sea4}). A condensate should be a (approximately) time-independent solution during inflation, hence upon substituting \eqref{LVsol} onto the evolution equation \eqref{anomevol}, we obtain that a (approximately)  constant solution $\tfrac{d}{dt} <\mathcal K^0 > \simeq 0$ necessitates a constant $H$ (inflation) and also the condition~\cite{bms1,bms2,ms1}:
\begin{align}\label{condition}
0 \simeq 1 - n_\star \, \frac{1.1}{3\,\pi^2} \, 
\Big(\frac{H}{M_{\rm Pl}}\Big)^2 \, \frac{\mu^4\, \mathcal A_{\rm CS}}{M_s^2\, M_{\rm Pl}}\,,    
\end{align}
during inflation. In the context of string theory \eqref{sea4}, for which the coefficient $\mathcal A_{\rm CS} $ is given by  $\frac{1}{96}\, \sqrt{\frac{2}{3}} \frac{M_{\rm Pl}}{M_s^2}$ ({\it cf.}  \eqref{genericCS}, \eqref{sea4}), 
this condition translates to:
\begin{align}\label{condition2}
1 \simeq  3 \times 10^{-4} \, n_\star \,\Big(\frac{H}{M_{\rm Pl}}\Big)^2 \, \Big(\frac{\mu}{M_s}\Big)^4\quad \Rightarrow \quad 
n_\star^{1/4} \,\frac{\mu}{M_s} \, \sim \,  7.6 \times \Big(\frac{M_{\rm Pl}}{H}\Big)^{1/2}\,.
\end{align}

\begin{exercise}
\begin{quotation}  Verify \eqref{condition} and \eqref{condition2}. \end{quotation} 
\end{exercise}

In \cite{bms2}, we have assumed $n_\star \simeq \mathcal O(1)$. In that case,  
upon taking the Planck data results for the upper bound of the inflationary Hubble scale $H_I$~\cite{planck}  
\begin{align}\label{HIPl}
\frac{H_I}{M_{\rm Pl}} \lesssim 10^{-5}\,, 
\end{align}
we obtain from Eq.~\eqref{condition2}, 
$\mu \gtrsim 2.4 \times 10^3 \, M_s$. In \cite{bms2,ms1,ms2} we assumed that the graviton modes are allowed to have momenta up to Planck scale, thus taking $\mu \sim M_{\rm Pl}$.\footnote{\label{ft5} In \cite{bms2,ms1} we followed the Green's function method of \cite{stephon}, instead of the Fourier method of \cite{lyth} adopted here, in order to evaluate the condensate. The two methods cannot be directly compared, especially in view of the various approximations involved. Nonetheless, as we see by comparing \eqref{condition2} with the corresponding one in \cite{bms2}, in our model, the two methods yield qualitatively similar results, that agree in order of magnitude, as expected for consistency.}  This determined the string scale at a high value and is consistent with the transplanckian conjecture, that no momenta of the effective field theory exceeds the Planck scale. 
However, from the point of view of a specific microscopic string theory, it may seem more appropriate to consider $\mu \sim M_s$, as mentioned above, and adopted in \cite{lyth}, but considering the proper number density of sources $n_\star$ as the adjustable parameter that will guarantee the formation of time-independent gravitational-anomaly condensates. In this scenario, one obtains from 
\eqref{condition2} and \eqref{HIPl}, 
\begin{align}\label{nstarval}
n_\star \gtrsim 3.3\times 10^{13}\,, 
\end{align} 
which defines the macroscopic number of sources (per proper volume) needed to produce a gravitational anomaly condensate in the context of an effective Chern-Simons gravitational theory, inspired from strings, with the string scale $M_s$ playing the r\^ole of the UV cutoff in the theory. In this second approach, the string scale is arbitrary and the conditions for the formation of the condensate translate into bounds on the number density of the sources of GW that lead to the condensate. 

\begin{important}{Important}
Thus, upon formation of the condensate during inflation ($H = H_I \simeq $ constant), we obtain a constant cosmic rate for the KR axion field, which we parametrise as~\cite{bms1}: 
\begin{align}\label{LVpar}
\dot b = \mathcal A_{\rm CS} \,  \mathcal K^0 \simeq {\rm constant} = \sqrt{2\epsilon} \, H \, M_{\rm Pl}.
\end{align}
The parameter  $\epsilon \ll 1$ needs to be compatible with the slow-roll cosmological data~\cite{planck} (see discussion below, Eq.~\eqref{freedman}).

This background solution {\it violates spontaneously} Lorentz symmetry. This can be readily seen from the duality relation 
\eqref{dual}, which connects $\dot b$ with the dual of the Kalb-Ramond torsion field, 
\begin{align}\label{LVspont}
{\rm constant} = \dot b \propto \epsilon^{ijk} \mathcal H_{ijk}, \, i,j,k=1,2,3. 
\end{align}
This implies the dynamical selection of a preferred Lorentz frame by the ground state of the theory, in which the spatial components of the totally antisymmetric torsion of the system is constant.  As we shall discuss later on, this property connects this gravitational theory with a SME effective field theory at the end of the inflationary period, when {\it chiral fermionic matter}, along with gauge fields, is assumed generated, according to the  approach of \cite{bms1,bms2,ms1,ms2}.
\end{important}

We now remark that the creation of the anomaly condensate \eqref{condensateN2} produces in principle a linear potential for the KR axion ({\it cf.} \eqref{sea4}) 
\begin{align}\label{linear}
V(b) = b (x)  \, \mathcal A_{\rm CS}\, \langle \, R_{\mu\nu\rho\sigma} \, \widetilde R^{\mu\nu\rho\sigma} \rangle_{\rm condensate \mathcal N}\, ,
\end{align}
with
\begin{align}\label{condval}
&\mathcal A_{\rm CS}\,  \langle \, R_{\mu\nu\rho\sigma} \, \widetilde R^{\mu\nu\rho\sigma} \rangle_{\rm condensate \mathcal N} 
\stackrel{\rm Eq.\eqref{LVpar}}{=} n_\star \, \frac{1.1}{\pi^2} \, \Big(\frac{H}{M_{\rm Pl}}\Big)^4 \, \mu^4\, \frac{\sqrt{2\epsilon} \, M_{\rm Pl}^2}{M_s^{2}}\,
\sqrt{\frac{2}{3}}\, \frac{M_{\rm Pl}}{96\, M_s^2}\, \nonumber \\
&\stackrel{\rm Eq.\eqref{HIPl}}{\lesssim} 4.3 \times 10^{-10} \, \sqrt{\epsilon} \, M_{\rm Pl}^3\,,
\end{align}
where in the last inequality we saturated the bound \eqref{nstarval}, for concreteness.

The potential \eqref{linear} 
is reminiscent of the linear axion-monodromy inflation potentials from appropriate brane compactifications (in, say, type IIB strings~\cite{silver}). Such linear potentials have been argued to lead to slow-roll hill-top inflation. However,  in our case the situation is very different. As we shall argue next, inflation in our scenario arises due to the non-linearities of the Running-Vacuum-Model  (RVM)-type vacuum energy, in particular the condensate induced $H^4$ term, without the need for external fields. The linear axion potential then serves merely as a consistency check of the slow-roll KR axion \eqref{LVsol} which characterises our case and leads to the parametrization \eqref{LVpar}. In \cite{bms1,bms2,ms1,ms2} we have taken $\epsilon \sim \mathcal O(10^{-2})$, as a generic 
slow-roll parameter of the cosmological data~\cite{planck}, but this is not restrictive, given that $b$ is not the inflaton, in the sense that it is not the linear potential of the KR axion that drives inflation in our case but the RVM non linearities. We shall discuss in the next subsection \ref{sec:rvm} the RVM properties of our inflation, and come back to this issue of estimating theoretically the order of magnitude of the phenomenological parameter $\epsilon$ (see Eq.~\eqref{freedman}). 

For the moment, we proceed to estimate the condensate contribution to the vacuum energy density.
Indeed, under the formation of a condensate \eqref{condensateN2}, one may expand the effective action \eqref{sea4} about this condensate, by writing for the gravitational Chern Simons term: 
\begin{align}\label{condDS}
b(x) \, R_{\mu\nu\rho\sigma} \, \widetilde R^{\mu\nu\rho\sigma} = \langle b(x)\, R_{\mu\nu\rho\sigma} \, \widetilde R^{\mu\nu\rho\sigma} \rangle_{\rm condensate \mathcal N} \, + \, : b(x)\, R_{\mu\nu\rho\sigma} \, \widetilde R^{\mu\nu\rho\sigma} :\,,
\end{align}
where $: \dots :$ denotes normal ordering ({\it i.e.} the creation operators appearing in the pertinent quantum-field correlation functions are placed on the  of the left of the annihilation operators), which ensures that the vacuum expectation value of the second term vanishes, upon quantization. The condensate term behaves as a de-Sitter type cosmological constant in the following sense~\cite{bms1,ms1}: the integrated solution of \eqref{LVpar}, implies $b(t)=\overline b(t_0) + \sqrt{2\epsilon} \, H \, (t-t_0) \, M_{\rm Pl}$, where $t_0$ denotes the beginning of inflation. The duration of inflation $\Delta t$ is given by $H \Delta t = N_e$, where $N_e = \mathcal O(60-70)$ is the number of e-foldings~\cite{planck}, thus in order for $b(t)$ not to change order of magnitude during the entire inflationary period,  we may require 
\begin{align}\label{bcb}
|\overline b(t_0)| \gtrsim N_e \, \sqrt{2\epsilon} \, M_{\rm Pl}\, = \mathcal O(10^2)\,\sqrt{\epsilon}\, M_{\rm Pl}\,,  
\end{align}
in which case the condensate term behaves approximately as a de Sitter (positive) cosmological constant term, provided (in our conventions) $\overline b(t_0) < 0$~\cite{bms1}. 

The total vacuum energy, with contributions from the KR axion ($b$) terms, the  gravitational Chern-Simons (non condensate) terms, proportional to the Cotton tensor \eqref{cottdef}, and the condensate itself (cond), can be obtained by the total stress energy tensor appearing in the appropriate Einstein equations \eqref{einsteincs}, upon inclusion of a de Sitter term. 
It can be easily shown (using \eqref{LVpar}) that the dominant term in the early Universe vacuum energy density is the one due to the condensate, which acquires the form
\begin{align}\label{condensatefinal}
\rho_{\rm cond} =  1.3 \times 10^{-3} \, \sqrt{\epsilon} \,  \frac{|\overline b(0)|}{M_{\rm Pl}}\, \Big(\frac{\mu}{M_s}\Big)^4 \, n_\star \, H^4  ,
\end{align}
where the various quantities appearing in \eqref{condensatefinal} have been defined previously. The reader should recall that consistency of our approach requires the condition \eqref{condition2}, which, on saturating the bounds \eqref{HIPl}, \eqref{nstarval}, for definiteness, yields ({\it cf.} \eqref{condval}):
\begin{align}\label{condestim}
\rho_{\rm cond} \equiv \frac{\Lambda}{\kappa^2} \, \sim 4.3 \times 10^{10} \, \sqrt{\epsilon}\, \frac{|\overline b(0)|}{M_{\rm Pl}} \, H^4 \,. 
\end{align}
Given \eqref{bcb}, this implies that the condensate term dominates over any other contributions to the vacuum energy coming from the KR axion field $b$ or the Cotton tensor due to the gravitational Chern-Simons term~\cite{bms1,bms2,ms1,ms2}. We leave the verification of this as an exercise to the reader.

\begin{exercise}\label{cottexer}
\begin{quotation}  Start from the property \eqref{cottprop} of the Cotton tensor, for the temporal component $\nu=0$, and 
replace the Chern-Simons anomaly term $R_{\alpha\beta\gamma\delta}\, \widetilde R^{\alpha\beta\gamma\delta}$ on the right-hand-side by its condensate. Consider the left-hand side of \eqref{cottprop} on a homogeneous and isotropic de Sitter background and thus argue, using also \eqref{tracecott}, that a constant $C_{00}$ arises as a consistent solution of this equation. Show that the constant $C_{00}$  is a negative quantity, but leads to subleading in magnitude contributions to the total vacuum energy compared to the condensate contributions. Also show that the energy density of the KR field, stemming from the stress tensor \eqref{axionst},  is subleading to the energy-density contribution of the condensate term (you should make use of the parametrisation \eqref{LVpar}, assuming simply $\epsilon \ll 1$). 
\end{quotation} 
\end{exercise}

We also leave it as an exercise to the reader to prove that the equation of state of this cosmological fluid satisfies~\cite{ms2} is of de-Sitter 
type during the condensate phase, in the sense that the total pressure ($p^{\rm total}$) and energy ($\rho^{\rm total}$) density, including KR axion ($b$), gravitational Chern-Simons (Cotton tensor, $C_{\mu\nu}$, gravitational anomaly (gGS)) contributions, and condensate \eqref{condestim} contributions, satisfy: 
\begin{align}\label{dseos}
p^b + p^{\rm gCS} = - (\rho^b + \rho^{\rm gCS})\, > \, 0, \quad {\rm and} \quad p^{\rm total} = - \rho^{\rm total}\, <\, 0\,,
\end{align}
where the superscript ``total'' denotes the (algebraic) sum of contributions from the $b$-axion, gCS and the (dominant) condensate $\Lambda$ terms.

\begin{exercise}
\begin{quotation} Consider the total (modified) stress-energy tensor in our string-inspired cosmology, 
\begin{align}\label{totstress}
T^{\rm total}_{\mu\nu}  = T^b_{\mu\nu} + \mathcal A_{\rm CS}\, C_{\mu\nu} + \Lambda g_{\mu\nu}.
\end{align}
Using the results of exercise \ref{cottexer}, on the estimate of the approximately constant $C_{00}$, as well as Eq.~\eqref{LVpar}, the trace property \eqref{tracecott} and 
the conservation equation \eqref{improved}, for an inflationary background spacetime, prove the following:
\begin{align}\label{eoscomp}
p^b&=\rho^b \quad ({\rm stiff~massless~axion~matter}) , \quad p_{\rm cond}^\Lambda=-\rho_{\rm cond}^\Lambda\, , \nonumber\\
p^{\rm gCS} &= \frac{1}{3} \rho^{\rm gCS}\, , \quad \rho^b = -\frac{2}{3} \rho^{\rm gCS}\,,
\end{align}
where the pressure density terms $p^{\rm gCS}$ are associated with the spatial diagonal components
of the Cotton tensor \eqref{cottdef}, $C_{ii}$, no sum over $i=1,2,3$, whilst the energy-density $\rho^{\rm gCS}$ is linked
to the temporal components $C_{00}$~\cite{bms1,ms2}.
From \eqref{eoscomp}, then, prove 
\begin{align}\label{eoscomp2}
\rho^b + \rho^{\rm gCS} &= \frac{1}{3} \rho^{\rm gCS} = -\frac{1}{2}\, \rho^b \, <\, 0 \,, \nonumber \\
p^b + p^{\rm gCS} &= -(\rho^b + \rho^{\rm gCS}) \, > \, 0\, 
\end{align}
thus proving the de-Sitter (RVM-type) equation of state \eqref{dseos}.
\end{quotation}
\end{exercise}

A remark is in  order regarding the contributions of the non-condensate anomaly terms in \eqref{dseos} (and \eqref{eoscomp2}), which are negative and such that the total energy density of the KR axion plus the Cotton-tensor-dependent anomaly terms is negative, satisfying though a de-Sitter-like equation of state.
Were it not for the condensate-$\Lambda$-\eqref{condestim} dominance, whose energy density is positive, the system would behave as an exotic one with ``phantom matter''~\cite{Sol\`aphantom1,Sol\`aphantom2}. 
The condensate dominance ensures that the vacuum of this string-inspired cosmology is characterised by a dominant {\it positive} vacuum energy of the form \eqref{condestim}, with a de-Sitter-like equation of state \eqref{dseos}.   
The reader should also observe from the result of the first line of \eqref{eoscomp2}, in combination with the estimate \eqref{LVpar}, that the total energy density of our cosmological fluid, including the condensate \eqref{condestim} reads~\cite{bms1,bms2,ms1}: 
\begin{align}\label{totalenerden}
\rho^{\rm total} = \rho^b + \rho^{\rm gCS} + \rho_{\rm condensate}^\Lambda = -\frac{1}{2}\, \epsilon \, M_{\rm Pl}^2\, H^2 + 4.3 \times 10^{10} \, \sqrt{\epsilon}\, \frac{|\overline b(0)|}{M_{\rm Pl}} \, H^4\,.
\end{align}
This form of the energy density is that of the running vacuum model (RVM) of Cosmology~\cite{Sol\`a,Sol\`a2,fossil,Sol\`alatest}, whose main features we review briefly in the next subsection \ref{sec:rvm} for completeness. This will also clarify the type of inflation induced by the condensate, since so far we have simply assumed a constant Hubble parameter to estimate the anomaly condensate, without specifying the microscopic origin of inflation.

As we shall show below, the inflation in our case is due to the non-linearities of the condensate term \eqref{condestim}, which depends on the fourth power of the Hubble parameter, and dominates in the early universe. No external inflaton fields are required. The KR axion field will provide though a slowly-moving pseudoscalar field during this RVM inflation~\cite{bms1,bms2,ms1,ms2}, whose rate of change can be constrained by the cosmological data~\cite{planck}.

\subsection{Condensates and Running-Vacuum-Model inflation \label{sec:rvm}}
 
The RVM cosmology~\cite{Sol\`a,Sol\`a2,fossil,Sol\`alatest} is an effective cosmological framework, with a cosmic-time-varying 
dark energy $\Lambda(t)$, which, nonetheless, is still characterised by an equation of state of de Sitter type:
\begin{align}\label{rvmeos}
p_{\rm RVM} (t)= -\rho_{\rm RVM}(t)
 \end{align}
 where $p$ ($\rho$) denotes the vacuum pressure (energy) density. The energy density is a function of even powers of the Hubble parameter $H(t)$ as a result of general covariance~\cite{Sol\`a,Sol\`a2,fossil,Sol\`alatest}:\footnote{The expression \eqref{rvmener} is the integrated form of the initially proposed `renormalization-group(RG)-like' evolution of the energy density, with $H(t)$ playing the r\^ole of the RG scale~\cite{Sol\`a,Sol\`a2,fossil}, 
 \begin{align}\label{rg}
 \frac{d}{d{\rm ln}H}\rho_{\rm RVM} = \sum_{i=1}^{\infty} c_i H^{2i} ,
 \end{align}
 with $c_i$ constant dimensionful in general coefficients (except the $c_4$ coefficient which is dimensionless in (3+1)-dimensions).
 In general, the expansion also includes terms $\dot H$, which however can be expressed in terms of $H^2$ and the deceleration parameter $q$. In most of the realistic applications, the various epochs of the Universe are characterised roughly by constant $q$'s and as such the expansion in even powers of $H^2$ suffices.} 
\begin{align}\label{rvmener}
\rho_{\rm RVM} (t) \equiv \frac{1}{\kappa^2} \Lambda(t) = \frac{3}{\kappa^2} \Big(c_0 + \nu \, H(t)^2 + \frac{\alpha}{H_I^2} \, H(t)^4 + \frac{\zeta}{H_I^4}  \, H(t)^6 + \dots \Big),
\end{align} 
in a standard parametrisation within the RVM framework, where $H_I$ is a fixed inflationary scale (obtained from the data, \eqref{HIPl}), and the $\dots$ denote higher powers of $H^2(t)$. The (dimensionless) coefficients $\nu, \alpha, \zeta, \dots$ can be determined either phenomenologically, by fitting the model with the data, especially at late eras, or can be computed within specific quantum field theory models~\cite{rvmqft1,rvmqft2,Sol\`alatest}.  The RVM framework provides a smooth cosmic evolution of the Universe~\cite{lima1,lima2}, explaining its thermodynamical and entropy production aspects, as a result of the decay of the running vacuum~\cite{therm1,therm2,therm3}, 
and a viable alternative to the $\Lambda$CDM at late epoch, with in-principle observable deviations, compatible with the current phenomenology~\cite{rvmpheno,rvmpheno2,rvmpheno3} (see also \cite{rvmpheno4,Tsiapi} for fits of general $\Lambda$-varying cosmologies). The RVM  framework also provides potential resolutions~\cite{rvmtens} to the recently observed, persisting tensions in the current-epoch cosmological data~\cite{tens1,tens2,tens3}, provided the latter do not admit mundane astrophysical and/or statistical explanations~\cite{mund}.

Phenomenologically, truncation of the expansion of the right-hand side of \eqref{rvmener} to terms of fourth power in $H(t)$ suffices to describe the entire Universe evolution from inflation at early epochs to the current era, where the r\^ole of the cosmological constant is played by the constant $c_0$, which appears as an integration constant when passing from the differential \eqref{rg} to the integrated form \eqref{rvmener} of the vacuum energy density, assuming, as is standard in RVM, that the entire evolution of the Universe is explained by \eqref{rvmener}, with constant coefficients $c_0, \nu, \alpha$. However, in the case of microscopic systems, such as the string-inspired one we discuss here, there may be phase transitions separating the various eras, and as a result the coefficients of the RVM evolution might change from era to era. 
 Moreover, within local quantum field theory studies~\cite{rvmqft1,rvmqft2,Sol\`alatest}, at least in non-minimally coupled scalar fields to gravity examined in those works, there is no coefficient $H^4$ arising, but only $H^2$ and $H^6$ (and higher). As we discussed in \cite{ms1} and review here, a term $H^4$ arises as a result exclusively of the condensation of gravitational anomalies in this string-inspired Chern-Simons modified theory. As we shall discuss below, the higher than $H^2$ non-linear terms provide inflation within the RVM framework, without the need for external inflaton fields,. 
 
Indeed, let us restrict our attention to the case \eqref{rvmener} (relevant for our purposes here). Let us denote collectively quantities referring to matter and radiation with the suffix ``m''. The pertinent equation
of state reads $p_m = \omega_m\, \rho_m$, which can be added to the RVM framework in such a way that the total energy and pressure densities, including the vacuum (RVM) contributions, are given by $p_{\rm total} = p_{\rm RVM} + p_m, \,\rho_{\rm total} = \rho_{\rm RVM} + \rho_m$. From the conservation of the total stress tensor of vacuum matter and radiation 
 one obtains the following evolution equation for the Hubble parameter $H(t)$~\cite{lima1,lima2}: 
 \begin {align}\label{flow}
\dot H + \frac{3}{2} \, (1 + \omega_m) \, H^2 \, \Big( 1 - \nu - \frac{c_0}{H^2} - \alpha \, \frac{H^2}{H_I^2} \Big) =0\,.
\end{align}
Ignoring $c_0$ (which, as we shall see, is a consistent assumption in our case), leads to a solution for $H(a)$ as a function of the scale factor $a$ (in units of the present-era scale factor) and the equation of state $\omega_m$ of  matter/radiation:
\begin{equation}\label{HS1}
 H(a)=\left(\frac{1-\nu}{\alpha}\right)^{1/2}\,\frac{H_{I}}{\sqrt{D\,a^{3(1-\nu)(1+\omega_m)}+1}}\,,
\end{equation}
where $D>0$ is an integration constant. For the early Universe, $ a \ll 1$, and thus one may assume without loss of generality that
$D\,a^{3(1-\nu)(1+\omega_m)} \ll 1$. On account of \eqref{HS1}, then, this leads to
an (unstable) dynamical early de Sitter phase, characterised by
an approximately constant Hubble parameter, $H_{\rm de~Sitter}  \simeq \left(\frac{1-\nu}{\alpha}\right)^{1/2}\,\, H_{I}$.

\begin{exercise} 
\begin{quotation} Starting from \eqref{flow}, and assuming $c_0=0$, prove that  a solution for $H(a(t))$ is given by \eqref{HS1}. \end{quotation}
\end{exercise}

It can be seen that at the current epoch, where $a(t) \gg 1$, and one has matter dominance ($\omega_m \simeq 0$), there are in principle observable deviations from the $
\Lambda$CDM model, still compatible though with the current phenomenology, due to the non-trivial $\nu H^2$ term in \eqref{rvmener} which dominates today. Phenomenologically, by fitting the CMB, weak- and strong- lensing, and baryon-acoustic-oscillation data~\cite{rvmpheno,rvmpheno2,rvmpheno3,rvmpheno4,Tsiapi}, one obtains $0 < \nu =\mathcal O(10^{-3})$ today, which incidentally is the order of magnitude of this parameter required by consistency of the RVM with Big-Bang-Nucleosynthesis (BBN) data~\cite{BBN}.

In our string-inspired model, as discussed in the previous subsection, we observe that the dominant condensate term \eqref{condestim} is of the RVM form \eqref{rvmener}, with the constant $\alpha \sim \sqrt{\epsilon} \, \frac{|\overline b(0)|}{M_{\rm Pl}} \sim \mathcal O(10^2) \epsilon $, if one saturates the bound \eqref{bcb}. In general, the total energy density of the vacuum
\eqref{totalenerden}
 is of RVM form, with $c_0=0$. The coefficient of the $H^2$ term, though, is negative, in contrast to the conventional RVM. This is due to the effects of the Chern-Simons (quadratic in curvature) terms in the effective action \eqref{sea4}.  In the microscopic model of of \cite{bms1}, during the post inflationary period, cosmic electromagnetic background fields can switch the sign of this term to a positive one, thus recovering the conventional RVM form at late epochs. This RVM form during inflation is consistent with the condensate itself inducing inflation at early epochs of the Universe evolution, according to the arguments leading to \eqref{HS1}.
 So the estimate of the condensate \eqref{condestim}, in a constant inflationary background with $H$ constant, is self consistent~\cite{bms1,bms2,ms1,ms2}.
 
 Let us now comment briefly on estimating the order of magnitude of $\epsilon$~\cite{ms2}. To this end, we may assume that the presence of the condensate is compatible with the Freedman equation for this Universe, implying that, during inflation, one has:
\begin{align}\label{freedman}
\frac{3}{\kappa^2} \, H^2  &= \rho^{\rm total} \simeq \rho_{\rm condensate}^\Lambda =  4.3 \times 10^{10} \sqrt{\epsilon} \frac{|\overline b(0)|}{M_{\rm Pl}} H^4 \, \, \nonumber \\
& \stackrel{\rm Eq.\eqref{condestim}}{\Rightarrow} \quad \epsilon \sim 7 \times 10^{-3} = \mathcal O(10^{-2})\,,
\end{align} 
where we saturated the bounds \eqref{bcb} and \eqref{HIPl}, for definiteness. The order of magnitude of $\epsilon$ is thus the same as the one assumed in \cite{bms1,ms2}. This should be considered as an allowed upper bound. On account of \eqref{bcb}, for $N_e = \mathcal O(60-70)$,  this value imply transplanckian values for the magnitude of $\overline b(0)$, $|\overline b(0)| \gtrsim \, 8.4 M_{\rm Pl}$. This does not affect the transplanckian conjecture, since the effective action depends only on $\dot b$ which assumes sub-planckian values. It may be in conflict though with the so-called distance conjecture of swampland~\cite{swamp}, which, however seems to affect also almost all single-field inflation models. This issue can only be resolved within the full UV complete string theory framework, and is beyond the effective field theory we are considering here, and beyond our purposes. Finally, we conclude this section by remarking that, with an $\epsilon=\mathcal O(10^{-2})$, the RVM coefficient $\alpha$ ({\it cf.} \eqref{rvmener}) in our model turns out to be of $\mathcal O(1)$, and positive, while the $\nu$ coefficient assumes the value $\nu = -\frac{1}{6}
\epsilon \sim  - 1.7 \times 10^{-3}$. Both coefficients are of the same order of magnitude 
as the corresponding ones in \cite{ms2}.\footnote{The alert reader might have noticed different numerical factors in front of the $H^4$ terms in the vacuum energy density \eqref{totalenerden}, as compared to those in \cite{bms1,bms2,ms1}. This is due to the fact that in estimating the condensate \eqref{condestim} we followed here the method and normalizations of \cite{lyth} instead of \cite{stephon}. However, as we have just seen, and already remarked in footnote \ref{ft5}, there are no qualitative or quantitative changes in the main phenomenological conclusions between these two frameworks.} 

We leave as a series of exercises for the reader to discuss the potentially drastic r\^ole of non-trivial, cosmic-time dependent, dilatons 
for the fate of the condensate, during RVM inflation, in a toy model.

\begin{exercise} 
\begin{quotation} Consider the string effective action in the presence of non-trivial dilatons $\Phi$, under the constraint \eqref{modbianchi2} in the absence of gauge fields, in its dual form, that is, the effective action written in terms of the (canonically normalised) Lagrange multiplier KR axion fields~\cite{kaloper2}:
\begin{align}\label{seab}
S^{\rm eff}_B & \simeq 
\; \int d^{4}x\, \sqrt{-g}\Big[ \dfrac{1}{2\kappa^{2}}\, R - \frac{1}{2\, \kappa^2} \partial_\mu \Phi \, \partial^\mu \Phi - \frac{1}{2}\, e^{-2\Phi} \, \partial_\mu b \, \partial^\mu + \mathcal A_{\rm CS} \,\partial_\mu b(x) \,\mathcal K^\mu\, 
+ \dots \Big], 
\end{align}
where $\mathcal K^\mu$ is the gravitational anomaly current \eqref{pontryaginA}, and $\mathcal A_{\rm CS} = \frac{1}{96}\, \sqrt{\frac{2}{3}} \frac{M_{\rm Pl}}{M_s^2}$ ({\it cf.} \eqref{genericCS}), with $M_s$ the string scale. 

\begin{enumerate}

\item[\textbf{(i)}]  You should notice that there is no dilaton coupling in the Chern-Simons anomaly term. Explain briefly this feature. 

\item[\textbf{(ii)}]  Consider a homogeneous and isotropic cosmological model based on the action \eqref{seab}, and 
assume that a {\it slowly-varying with the cosmic time} condensate for the Chern-Simons gravitational anomaly has been formed, 
so that \eqref{condensateN2} is in operation, but in an RVM form, {\it i.e.} one should replace the constant $H_I$ by a slowly varying $H(t)$:
\begin{align}\label{condHt}
\langle R_{\mu\nu\rho\sigma}\, \widetilde R^{\mu\nu\rho\sigma} \rangle_{\rm condensate} &\simeq n_\star \, \frac{1.1}{\pi^2} \, 
\Big(\frac{H(t)}{M_{\rm Pl}}\Big)^3 \, \mu^4\, \frac{\dot b(t)}{M_s^{2}} \nonumber \\
& \simeq 3.5 \times 10^{12} \, \Big(\frac{H(t)}{M_{\rm Pl}}\Big)^3\, M_s^2 \dot b(t) 
~\,,
\end{align}
where in the second (approximate) equality (which you should verify explicitly) we used, for definiteness, an $n_\star$ that satisfies \eqref{condition2}
for a constant inflationary scale $H_I$  saturating the upper bound of \eqref{HIPl}, as inferred from the data~\cite{planck}. $H_I$ here should not be identified with $H(t)$. This would ensure that in the absence of non trivial dilatons, one would recover the situation discussed previously, with $H\simeq H_I$.
 
\item[\textbf{(iii)}]  Write down the dilaton and KR axion equations of motion, derived from the homogeneous and isotropic cosmological version of the action \eqref{seab}.

\item[\textbf{(iv)}]  Show that 
\begin{align}\label{bphiconst}
 e^{-2\Phi(t)} \dot b =   \mathcal A_{\rm CS}\, \mathcal K^0\,,
 \end{align} 
 is a solution of the KR axion equation of motion.

 \item[\textbf{(v)}]  Make the assumption that $\dot b$ in \eqref{condHt} can be replaced by the one in the solution \eqref{bphiconst} in terms of the dilaton $\Phi(t)$. Then, by approximating the anomaly-current equation in  a Robertson-Walker background corresponding to $H(t)$ as
 \begin{align}\label{ancurrcondHt}
 \langle \nabla_\mu \mathcal K^\mu \rangle  
 \simeq \frac{d}{dt} \langle \mathcal K^0 \rangle + 3 \,H(t) \, \langle \mathcal K^0\rangle  \simeq \langle R_{\mu\nu\rho\sigma}\, \widetilde R^{\mu\nu\rho\sigma} \rangle_{\rm condensate},
 \end{align}
derive the condition for the validity of \eqref{ancurrcondHt}:
\begin{align}\label{Hphi}
H(t) \, e^{\Phi(t)} \simeq 10^{-5} M_{\rm Pl} \,,
\end{align}
for an approximately time-$t$-independent condensate $\langle \mathcal K^0\rangle$.

\item[\textbf{(vi)}]  
Using \eqref{Hphi}, show that the dilaton equation of motion stemming from the action \eqref{seab} leads to: 
\begin{align}\label{dil2}
3 H^2 \, \dot H - (\dot H)^2 + H\, \ddot H \simeq 7.2 \times 10^{-15} \, \frac{M_{\rm Pl}^2}{M_s^4}\, \langle \mathcal K^0\rangle^2
\,.
\end{align}
\item[\textbf{(vii)}]  
In this toy cosmological model, assume the validity of \eqref{dil2} from an intitial cosmic time $t_0$ in which $H(t_0) \to H_I$, where 
$H_I$ saturates the observational bound  \eqref{HIPl}.
Then, assuming slow roll for $H$, in which $H\, \ddot H$ and $(\dot H)^2$ terms in \eqref{dil2} are subleading, show that 
one obtains an inflationary scenario with a higher-than-simple-exponential expansion for the scale factor $a(t)$ of this dilaton-dominated Universe, 
that is, show that 
\begin{align}\label{scalef}
a(t) \sim \exp\Big(\frac{3H_I}{4c_1} \Big[ \Big(1 + c_1 (t-t_0)\Big)^{4/3} -1 \Big]\Big)\,,
\end{align}
in units of $a(t_0)$, and determine the constant $c_1>0$.
Interpret the boundary condition $H(t_0)=H_I$ in the context of the model of \cite{ms1} discussed previously in this work.

\item[\textbf{(viii)}]  
Check and discuss the self consistency of the slow-roll assumption for $H(t)$ in part {\bf (vii)}.

\item[\textbf{(ix)}]  
Finally, by assuming the validity of the Friedmann equation, provide an estimate of $\langle \mathcal K^0\rangle$ for this RVM universe, in which the condensate \eqref{condHt} dominates the total energy density. To answer this part of the question, first discuss the conditions under which the background axion field $b(t)$, satisfying \eqref{bphiconst}, does not change order of magnitude under the entire duration of inflation.

\end{enumerate} 
\end{quotation}
\end{exercise} 

The above exercise on a potential r\^ole of the dilaton on the induced RVM inflation does not constitute a complete treatment within string theory. In the case of bosonic (or Heterotic) strings, in addition to the anomaly four-derivative (order $\alpha^\prime$) Chern-Simons term in the effective 
action, there are also four-derivative (quadratic in curvature) Gauss-Bonnet (GB) terms~\cite{amb1,amb2,amb3,amb4}, which are non-trivial when the dilaton is non-trivial. The exception is the type IIB string, for which the GB terms are absent. 

\begin{exercise}\label{ex:dil}
\begin{quotation}
Consider the string-inspired effective action \eqref{seab}, but in the presence of a quintessence-type dilaton-$\Phi$ potential, arising, for instance, in non-critical string cosmologies~\cite{aben,emn}: $V(\Phi)= \mathcal C\,  \exp(c_1\Phi)$, where $\mathcal C, c_1  \in \mathbb R$ are appropriate real constants. Determine $c_1$ and $\mathcal C$ such that a dilaton $\Phi=0$ is a consistent solution of the equations of motion, corresponding to the case studied in \cite{bms1,bms2,ms1,ms2}, in which an anomaly condensate is formed, with $\dot b = \mathcal A_{\rm CS} \langle \mathcal K^0 \rangle = {\rm constant}$. Comparing your results with the studies in \cite{aben,emn}, and using their definitions for super (sub) critical string, depending on the sign of $\mathcal C$, determine which type of non-critical string this situation corresponds to.
\end{quotation}
\end{exercise}

\section{Links with the Lorentz- and CPT- Violating Standard Model Extension, and leptogenesis}

We now come to the final, but also crucial, topic of our discussion, namely how the above results are linked to the Standard Model Extension~\cite{sme,sme2} with Lorentz and CPT Violation. This becomes possible if we consider the generation of fermionic (chiral) matter, which in our model occurs towards the end of the RVM inflationary period, as a consequence of the decay of the running vacuum. 

As discussed in \cite{bms1,bms2,ms1,ms2}, in the context of the precursor string-theory model, (chiral) fermionic matter, represented by a generic fermion $\psi$ for our purposes, for brevity, will couple to the (totally antisymmetric) torsion $\mathcal H_{\mu\nu\rho}$ via the gravitational covariant derivative. Adding the fermion action to the string action \eqref{sea},  
with $\Phi=0$, and performing the path-integration over the torsion field $\mathcal H_{\mu\nu\rho}$ 
in a curved background, with the $\delta$-functional constraint \eqref{modbianchi2}, implemented as in Exercise \ref{bianchconstr},
and being represented in terms of the pseudoscalar Lagrange multiplier field $b(x)$ (KR axion), 
one obtains the effective action:
 \begin{align}\label{sea6}
&S^{\rm eff} =\; \int d^{4}x\sqrt{-g}\Big[ -\dfrac{1}{2\kappa^{2}}\, R + \frac{1}{2}\, \partial_\mu b \, \partial^\mu b -  \sqrt{\frac{2}{3}}\,
\frac{\alpha^\prime}{96\, \kappa} \, \partial_\mu b(x) \, {\mathcal K}^\mu
\Big]  \nonumber \\
&+ S_{\rm Dirac~or~Majorana}^{Free} + \int d^{4}x\sqrt{-g}\, \Big[\Big( {\mathcal F}_\mu + \frac{\alpha^\prime}{2\, \kappa} \, \sqrt{\frac{3}{2}} \, \partial_{\mu}b \Big)\, J^{5\mu}    - \dfrac{3\alpha^{\prime\, 2}}{16 \, \kappa^2}\,J^{5}_{\mu}J^{5\mu} \Big] + \dots,
\end{align}
where $J^{5 \mu}~\equiv~\sum \, \overline \psi \gamma^5 \, \gamma^\mu \, \psi $ denotes 
the axial fermion current , ${\mathcal F}^d  =   \varepsilon^{abcd} \, e_{b\lambda} \,  \partial_a \, e^\lambda_{\,\,c}$, with $e^\mu_{\,\,c}$ the vielbeins (with Latin indices pertaining to the tangent space of the space-time manifold at a given point, in a standard notation), 
$S_{\rm Dirac~or~Majorana}^{Free}$ denotes the free-fermion kinetic terms, 
and the $\dots$ in (\ref{sea6}) indicate gauge field kinetic terms, as well as terms of higher order in derivatives. The action \eqref{sea6} is valid for both Dirac or Majorana fermions.\footnote{In case of multifermion theories, as reguired in phenomenologically realistic models, one simply has to sum the appropriate effective action terms over all the fermion species, with the 
axial current reading as $J^{5\mu}= \sum_{i=\rm fermion~species} \overline \psi_i \gamma^5 \gamma^\mu \psi_i$.} The reader is invited to take note of the presence in \eqref{sea6} of the CP-violating interactions of the derivative of the field $b$ with the axial fermion current $J^{5\mu} $, as well as of the repulsive axial-fermion-current-current term, 
$-\dfrac{3\alpha^{\prime\, 2}}{16 \, \kappa^2}\,J^{5}_{\mu}J^{5\mu}$, which is characteristic of theories with Einstein-Cartan torsion~\cite{torsion,torsshap}, as is our string-inspired model~\cite{kaloper}.  The proof of \eqref{sea6} is left as a set of exercises for the reader.

\begin{exercise}
\begin{quotation} Consider for definiteness a Dirac fermion $\psi$ in a curved space-time with a string-inspired totally antisymmetric torsion $\mathcal H_{\mu\nu\rho}$, as in \eqref{torsionconn}. On using the definition of the gravitational covariant derivative acting on the fermions in terms of vielbeins $e^a_{\,\,\mu}$ and the torsionful spin connection $\overline \omega_{\mu\,\,b}^{a}$ corresponding to \eqref{torsionconn} (where Latin indices are tangent-space indices), as well as properties of products of three Dirac $\gamma^\mu$ matrices, prove first that there is a linear coupling of the fermion axial current $\overline \psi \, \gamma^5 \, \gamma^\mu \, \psi$ 
to $\varepsilon_{\mu\nu\rho\sigma}\, \mathcal H^{\nu\rho\sigma}$, where the covariant Levi-Civita tensor density $\varepsilon_{\mu\nu\rho\sigma}$ has been defined in \eqref{coveps},  and determine its coefficient. Then, by adding this fermion action to \eqref{sea}, consider the constrained path-integration 
over $\mathcal H_{\mu\nu\rho}$, using a $\delta$-functional constraint for \eqref{modbianchi2}, as in Exercise \ref{bianchconstr}. On representing the $\delta$-functional by means of the canonically normalised Lagrange multiplier field $b(x)$, then, prove \eqref{sea6}.  Show also that, for Robertson-Walker backgrounds, in the absence of perturbations, the quantity ${\mathcal F}^d  =   \varepsilon^{abcd} \, e_{b\lambda} \,  \partial_a \, e^\lambda_{\,\,c}$ vanishes. 
\end{quotation}
\end{exercise} 

We next observe, that, in case of the spontaneous LV background \eqref{LVpar}, due to the anomaly condensate  in our cosmology, the fermion-axial-current-KR-axion interaction in \eqref{sea} leads to a LV and CPTV interaction with the background, which is of a SME type \eqref{smef}, with $a_\mu=0$ and 
\begin{align}\label{background}
b_\mu = M_{\rm Pl}^{-1} \, \dot{\overline b} \, \delta_{\mu\,0}\,, \quad  \mu=0, \dots 3\,, \quad \dot{\overline b} = {\rm constant}\,,
\end{align}
having only a temporal component, with $\overline{b}$ 
the solution to the KR equation of motion stemming from \eqref{sea4}.

\begin{important}{Important}
In the presence of massive right-handed neutrinos (RHN), with standard portals, coupling the RHN sector to Standard Model (SM) lepton and Higgs sectors, then, one can consider the following fermion action in the background \eqref{background}:
\begin{align}\label{smelag}
\mathcal{L}= {\mathcal L}_{\rm SM} + i\overline{N}\, \gamma^\mu\, \partial_\mu \, N-\frac{m_N}{2}(\overline{N^{c}}N+\overline{N}N^{c})-\overline{N}\gamma^\mu\, b_\mu \, \gamma^{5}N-\sum_f \, y_{f}\overline{L}_{f}\tilde{\phi}^dN+ {\rm h.c.}
\end{align}
where h.c.  denotes hermitian conjugate, ${\mathcal L}_{\rm SM}$ denotes the SM Lagrangian,
$N$ is the RHN field (with $N^c$  its charge conjugate field), of (Majorana) mass $m_N$,  $\tilde \phi$ is the SU(2) adjoint of the Higgs field  $\phi$ ($\tilde{\phi}^d_i \equiv \varepsilon_{ij}\phi_j~, \, i,j=1,2,$ SU(2) indices),
 and $L_{f}$ is a lepton (doublet) field of the SM sector, with $f$ a generation index, $f=e, \mu, \tau$, in a standard notation for the three SM generations; $y_f$ is a Yukawa coupling, which is non-zero and provides a non-trivial (``Higgs portal'') interaction between the RHN and the SM sector, used in the seesaw mechanism for generation of SM neutrino masses. 
As discussed in \cite{sarkar1,sarkar2,sarkar3,sarkar4}, and we shall describe briefly below, 
such backgrounds can produce phenomenologically correct leptogenesis. In particular, we consider lepton-number asymmetry originating from {\it tree-level} decays of heavy sterile RHN  into SM leptons.

Indeed, in the context of the model \eqref{smelag}, a lepton asymmetry is generated due to the CPV and CPTV tree-level decays of the RHN $N$ into SM leptons, in the presence of the background \eqref{background}, through ${\rm Channel} ~I: \,  N \rightarrow l^{-}h^{+}~, ~ \nu \, h^{0}$,  and 
${\rm Channel ~II}: \, N \rightarrow l^{+}h^{-}~,~  \overline \nu \, h^{0}$,
where $\ell^\pm$ are charged leptons, $\nu$ ($\overline \nu$) are light, "active", neutrinos (antineutrinos) in the SM sector, $h^0$ is the neutral Higgs field, and
 $h^\pm$ are the charged Higgs fields, which, at high temperatures, above the spontaneous electroweak symmetry breaking, of interest in this scenario~\cite{sarkar1,sarkar2,sarkar3,sarkar4,ms1}, do not decouple from the physical spectrum.  As a result of the non-trivial $b_0 \ne 0$ background (\ref{background}), the decay rates of the Majorana RHN between the channels I and II are different, resulting in a Lepton-number asymmetry.  
\end{important}

\begin{exercise} 
\begin{quotation} Consider the tree-level decays of a massive Majorana neutrino $N$, of mass $m_N$,  in the theory \eqref{smelag} into charged lepton and Higgs particles and antiparticles only, in the background \eqref{background}. 
By following standard particle physics methods, prove that the tree-level decay rates $\Gamma$, $N \to \ell^-h^+ $ and $N \to \ell^+h^-$, are given, respectively, by:
\begin{align}\label{lepto}
\Gamma_{N \to \ell^-h^+} &= \sum_{f=e,\mu,\tau} \frac{|y_f|^2}{32\, \pi^2} \, \frac{m^2_N}{\Omega}\frac{\Omega + b_0}{\Omega-b_0}, \quad  \Gamma_{N \to \ell^+h^-} = \sum_{f=e,\mu,\tau} \frac{|y_f|^2}{32\, \pi^2} \, \frac{m^2_N}{\Omega}\frac{\Omega - b_0}{\Omega + b_0},  \nonumber \\ {\rm with}\quad\Omega &= \sqrt{m_N^2 + b_0^2}\,.
\end{align}
 The reader should observe that the difference between these two rates vanishes for vanishing background $b_0 \to 0$.  To linear order in $b_0$, with $|b_0| \ll m_N$, argue that these decay rates may be interpreted as implying the presence of  ``effective'' RHN masses $m_{N \, \rm eff}^\pm = m_N \pm 2b_0$ in the $\ell^\mp \, h^\pm$ decay channels, respectively. 
 \end{quotation}
\end{exercise}

Such asymmetries in the decay rates produce lepton asymmetry, which can then be communicated~\cite{sarkar4} to the baryon sector 
by means of appropriate baryon(B) and lepton(L)-number violating but B-L conserving processes, e.g. sphalerons in the SM sector of the model~\cite{shap1,shap2,shap3,gavel1,gavel2}, according to standard leptogenesis scenarios~\cite{leptofuya,leptogenesis}.

Before closing, we remark that in the actual model of \cite{bms1,bms2,ms1,ms2} the situation is a bit more complicated than the simplified scenario with a constant background \eqref{background} surviving in the radiation phase. The generation of chiral fermions at the end phase of the RVM inflation in the model leads to a cancellation of the primordial gravitational anomalies by 
the ones generated by the chiral fermions themselves, leaving only possible chiral anomalies (in the gauge sector) surviving in the post-inflationary period. This leads, as a consequence, to a temperature dependence for the background $b_0 \propto T^3$, during the post inflationary period, as explained in detail in \cite{bms1}. For the short period of leptogenesis, such temperature dependent backgrounds are almost constant, and the 
resulting lepton asymmetry can be calculated analytically~\cite{sarkar1,sarkar2,sarkar4}, leading to similar, qualitatively and quantitatively, conclusions as the simple constant-background \eqref{background} case, reviewed above.

The advantage of the $T^3$ temperature dependence of the axion background $B_0$, which survives the inflationary period, is that one can trace it to the current era (up to complications including chiral anomalies, which can change the $T^3$ behaviour, {\it e.g.} to $T^2$, as discussed in some detail in \cite{bms1}). The current KR axion background is well below the current bounds \eqref{coeff} of this background~\cite{smebounds}. Specifically one finds~\cite{bms1} $b_0|_{\rm today} \sim 10^{-44}$~eV, if chiral anomalies are ignored (i.e., $T^3$ scaling), and  $b_0|_{\rm today} \sim 10^{-34}$~eV, if chiral anomalies take over at late epochs.
Even if one takes into account the relative motion of our Earthly laboratory frames with respect to the CMB frame (with velocity $v_i$, $|\vec v| = \mathcal O(390 \pm 60)$ km/sec~\cite{planck,anis} ), which leads to spatial components of the background $b_i = \gamma \frac{v_i}{c}\, b_0$, with $\gamma \sim 1$ the Lorentz factor,  the resulting spatial components $b_i$ of the LV and CPTV background lie comfortably within the existing bounds \eqref{coeff}~\cite{smebounds}.

The following exercise provides the reader with a simple way to understand the $T^3$ scaling of $\dot b$ in the absence of chiral anomalies, during the post-inflationary era of the cosmological model of \cite{bms1,bms2,ms1,ms2}. 

\begin{exercise}
\begin{quotation}
Consider the KR axion $b(x)$ equation of motion stemming from the effective action \eqref{sea6} for the case of a conserved axial-fermion current $J^{5\, \mu}(x)$ in the absence of gravitational anomalies ({\it i.e.} set $\mathcal K^\mu=0$).
\begin{enumerate}
\item[\textbf{(i)}] Show, that in a homogeneous and isotropic Robertson-Walker background, in the radiation-dominated era, there is a $T^3$ scaling 
of the cosmic rate of the $b(x)$ field, $\dot b$, where $T$ is the cosmic temperature (use standard cosmology arguments~\cite{kolb} to relate the scale factor of the Universe to the cosmic temperature $T$). 

\item[\textbf{(ii)}] By assuming~\cite{bms1} that the radiation era succeeds the inflationary one, during which the inflationary scale $H_I$ is related to the (de-Sitter observer dependent) Gibbons-Hawking temperature~\cite{gh}, $T =H_I/2\pi$, determine the $T$-scaling proportionality constant in the expression for $\dot b$ of part {\bf (i)}, using the parametrization \eqref{LVpar} (with an $\epsilon = \mathcal O(10^{-2})$, and a $H_I$ saturating the bound \eqref{HIPl}) at the exit phase of inflation/beginning of the radiation era in the framework of the model of \cite{bms1,bms2,ms1,ms2}. Thus show that this $\dot b$ satisfies the current LV and CPTV bounds \eqref{coeff}.
\end{enumerate} 
\end{quotation}
\end{exercise}

\section{Summary and oultook} 

With the above remarks  we conclude our discussion on how one can obtain Lorentz and/or CPT Violating terms, that appear phenomenologically 
in the SME effective Lagrangian, starting from a microscopic quantum gravity theory. 
Within our specific string-inspired cosmological field theory example, we have seen how 
condensates of primordial gravitational waves, of quantum-gravitational origin, can lead to spontaneous violation of Lorentz and 
CPT symmetries in low-energy effective theories, which contain terms of a form appearing in  SME Lagrangians. 

We have pointed out the crucial r\^ole of the UV complete theory of quantum gravity in leading to these condensates. The lack, however, of a complete understanding of energy regimes above the Planck scale, even in the context of UV complete theories, such as strings, 
has some consequences for the accuracy of the relative estimates. Nonetheless, we hope we made it clear to the reader that LV and CPTV processes might play a crucial r\^ole on the existence of our Universe, and thus ourselves, given, the potential link of the spontaneous violation of Lorentz and CPT symmetries by the condensates to the matter-antimatter asymmetry in our Universe, as we discussed above. An important aspect of our considerations is that the matter-antimatter asymmetry in the Cosmos might have a geometric origin, as a consequence of the close connection of LV to condensates of torsion (axion-like) fields, which characterise the massless gravitational multiplet of string theory (which is also the ground state of the phenomenologically relevant superstrings). 

From a phenomenological point of view it would be interesting to explore further the profile of the primordial GW generated during our RVM inflation, as well as the densities of primordial black holes during that era, especially in models with non-trivial dilatons, such as \eqref{seab} (in case of type-IIB-string inspired models), or extensions thereof, including Gauss-Bonnet-dilaton coupled combinations (in case of heterotic and bosonic strings). There is the possibility of enhanced gravitational perturbations and densities of primordial black holes in such models, which could affect the aforementioned GW profiles at post inflationary (radiation) eras, thus leading to observable in principle effects in interferometers. In addition, effects of the LV and CPTV SME-type background coefficients $b_\mu$ \eqref{background}, which are linked to forbidden atomic transitions and other modifications of atomic spectra~\cite{antiHcpt2,sme,lehnert,sme3}, might affect  BBN physics, given the increasing nature of this coefficient with the cosmic temperature, which might lead to further constraints. These are issues to be examined in the future, by extending, for instance, the LV analysis of \cite{Lambiase} to incorporate appropriately the CPTV effects arising in our framework.

\begin{acknowledgement} 
The author is grateful to Prof. C. L\"ammerzahl and Dr. C. Pfeifer for the invitation to contribute a chapter to this book
on {\it Modified and Quantum Gravity - From theory to experimental searches on all scales - WEH 740}, based on an invited 
talk in the 740 Wilhelm-und-Else-Heraeus-Seminar, 1-5 February 2021. This work  was supported partly by STFC (UK) Grant ST/T000759/1. NEM also acknowledges participation in the COST Association Action CA18108 ``{\it Quantum Gravity Phenomenology in the Multimessenger Approach (QG-MM)}''.
\end{acknowledgement}

 \end{document}